\documentclass[final,3p,times]{elsarticle}
\usepackage{Paper}
\journal{}
\newcommand{\RR}{\ensuremath{\mathbb{R}}}

\newcommand{\fluiddomain}{\ensuremath{\Omega}} 
\newcommand{\subdomain}{\ensuremath{\Omega_{0}}} 
\newcommand{\body}{\ensuremath{\Gamma}} 
\newcommand{\rigbodyi}{\ensuremath{\Gamma_{r}^{i}}}
\newcommand{\torsbodyi}{\ensuremath{\Gamma_{t}^{i}}}
\newcommand{\defbodyi}{\ensuremath{\Gamma_{d}^{i}}}
\newcommand{\rigidbody}{\ensuremath{\Gamma_r}} 
\newcommand{\torsbody}{\ensuremath{\Gamma_t}} 
\newcommand{\defbody}{\ensuremath{\Gamma_d}} 
\newcommand{\flowpoint}{\ensuremath{\bm{x}}} 
\newcommand{\sdpoint}{\ensuremath{\bm{x}^{0}}} 
\newcommand{\surface}{\ensuremath{s}} 
\newcommand{\bodypoint}{\ensuremath{\bm{\chi}}} 
\newcommand{\rigbodypoint}{\ensuremath{\bm{\chi}_{r}^{i}}}
\newcommand{\defbodypoint}{\ensuremath{\bm{\chi}_{d}^{i}}}
\newcommand{\torsbodypoint}{\ensuremath{\bm{\chi}_{t}^{i}}}
\newcommand{\defbodypointi}[1]{\ensuremath{\chi_{d{#1}}^{i}}} 
\newcommand{\hinge}{\ensuremath{\bm{\chi}^{0i}_{t}}} 
\newcommand{\velocity}{\ensuremath{\bm{u}}} 
\newcommand{\timet}{\ensuremath{t}} 
\newcommand{\pressure}{\ensuremath{p}} 
 
\newcommand{\reynolds}{\ensuremath{Re}} 
\newcommand{\stress}{\ensuremath{\bm{f}}} 
 
\newcommand{\discretedelta}{\ensuremath{d}} 
\newcommand{\discretedeltap}{\ensuremath{d_{p}}} 

\newcommand{\nbr}{\ensuremath{m_r}} 
\newcommand{\nbt}{\ensuremath{m_t}} 
\newcommand{\nbd}{\ensuremath{m_d}} 

\newcommand{\deflection}{\ensuremath{\theta}}
\newcommand{\deflectiondot}{\ensuremath{\phi}} 
\newcommand{\deflectiondoti}{\ensuremath{\phi^{i}}} 
\newcommand{\deflectioni}{\ensuremath{\deflection^{i}}} 
\newcommand{\deflectionoi}{\ensuremath{\deflection^{0i}}} 
\newcommand{\inertia}{\ensuremath{i_t^i}} 
\newcommand{\stiffness}{\ensuremath{k_t^i}} 
\newcommand{\damper}{\ensuremath{c_t^i}} 
\newcommand{\dynamics}{\ensuremath{g}_{t}} 
\newcommand{\inertiadim}{\ensuremath{I_t^i}} 
\newcommand{\stiffnessdim}{\ensuremath{K_t^i}} 
\newcommand{\damperdim}{\ensuremath{C_t^i}} 

\newcommand{\cauchystress}{\ensuremath{\bm{\sigma}^i}} 
\newcommand{\dynamicsdef}{\ensuremath{\bm{g}_{d}}} 
 
\newcommand{\defbodypointdoti}[1]{\ensuremath{\zeta_{d{#1}}^{i}}} 

\newcommand{\lengthscale}{\ensuremath{L}} 
\newcommand{\velocityscale}{\ensuremath{U_{\infty}}} 
\newcommand{\densityf}{\ensuremath{\rho_{f}}}
\newcommand{\densitys}{\ensuremath{\rho_{d}^i}}
\newcommand{\visc}{\ensuremath{\nu}} 
\newcommand{\factor}{\ensuremath{\alpha}} 

\newcommand{\streamfnc}{\ensuremath{s}} 
\newcommand{\velocityq}{\ensuremath{q}} 
\newcommand{\C}{\ensuremath{C}} 
\newcommand{\Ct}{\ensuremath{C^{T}}} 
\newcommand{\A}{\ensuremath{A}} 
\newcommand{\B}{\ensuremath{B}} 
\newcommand{\I}{\ensuremath{I}} 
\newcommand{\LL}{\ensuremath{L}} 
\newcommand{\Enk}[2]{\ensuremath{E_{#1}^{(#2)}}} 
\newcommand{\Etnk}[2]{\ensuremath{E_{#1}^{(#2)T}}} 
\newcommand{\En}[1]{\ensuremath{E_{#1}}} 
\newcommand{\Etn}[1]{\ensuremath{E_{#1}^{T}}} 
\newcommand{\Snk}[2]{\ensuremath{S_{t,#1}^{(#2)}}} 
\newcommand{\Stnk}[2]{\ensuremath{S_{t,#1}^{(#2)T}}}
\newcommand{\Sdnk}[2]{\ensuremath{S_{d,#1}^{(#2)}}}

\newcommand{\stressnk}[2]{\ensuremath{f_{#1}^{(#2)}}}
\newcommand{\stressn}[1]{\ensuremath{f_{#1}}}
\newcommand{\stressnormali}{\ensuremath{f_N^i}}
\newcommand{\dt}{\ensuremath{\Delta t}}  
\newcommand{\dx}{\ensuremath{\Delta x}}  
\newcommand{\ds}{\ensuremath{\Delta s}}  
\newcommand{\dsd}{\ensuremath{\Delta x^{0}}}  
\newcommand{\J}{\ensuremath{J_t}} 
\newcommand{\Jd}{\ensuremath{J_d^{(k)}}} 
\newcommand{\Ri}[2]{\ensuremath{R^{i{#2}}_{t{#1}}}} 
\newcommand{\Qi}[1]{\ensuremath{Q^{i{#1}}_{t}}} 
\newcommand{\R}[2]{\ensuremath{R_{t{#1}}^{#2}}} 
\newcommand{\Q}[1]{\ensuremath{Q_{t}^{#1}}} 
\newcommand{\Mdi}{\ensuremath{M_{d}^{i}}} 
\newcommand{\Rdi}{\ensuremath{R_{d}^{i}}} 
\newcommand{\Qdi}{\ensuremath{Q_{d}^{i}}}
\newcommand{\Wdi}[2]{\ensuremath{W_{d{#1}}^{i{#2}}}}
\newcommand{\Wd}[2]{\ensuremath{W_{d{#1}}^{#2}}} 
\newcommand{\Qd}[1]{\ensuremath{Q_{d}^{#1}}} 
\newcommand{\deflectionincrease}{\ensuremath{\Delta \deflection}} 
\newcommand{\deflectionnk}[2]{\ensuremath{\deflection_{#1}^{(#2)}}}

\newcommand{\deflectiondotnk}[2]{\ensuremath{\deflectiondot_{#1}^{(#2)}}}
\newcommand{\deflectiondotincrease}{\ensuremath{\Delta \deflectiondot}} 
\newcommand{\stresst}[1]{\ensuremath{f_{t,{#1}}^{i}}}
\newcommand{\stressd}[1]{\ensuremath{f_{d{#1}}^{i}}}

\newcommand{\defbodypointincrease}{\ensuremath{\Delta \chi_d}} 
\newcommand{\defbodypointdotincrease}{\ensuremath{\Delta \zeta_d}}
\newcommand{\defbodypointnk}[2]{\ensuremath{\chi_{d,#1}^{(#2)}}}
\newcommand{\defbodypointdotnk}[2]{\ensuremath{\zeta_{d,#1}^{(#2)}}}

\newcommand{\Pnk}[2]{\ensuremath{P_{#1}^{(#2)}}} 
\newcommand{\Ptnk}[2]{\ensuremath{P_{#1}^{(#2)T}}} 
\newcommand{\Eo}{\ensuremath{E_{0}}} 
\newcommand{\Eot}{\ensuremath{E_{0}^{T}}} 
\newcommand{\Eoij}[1]{\ensuremath{E_{0 #1}}} 
\newcommand{\Pn}[1]{\ensuremath{P_{#1}}} 

\newcommand{\fluidrhs}{\ensuremath{r^{f}_{n}}} 
\newcommand{\bcrhs}[1]{\ensuremath{r^{c(#1)}}} 
\newcommand{\deflectionrhs}[1]{\ensuremath{r^{\deflection(#1)}}} 
\newcommand{\deflectiondotrhs}[1]{\ensuremath{r^{\deflectiondot(#1)}}} 
\newcommand{\defbodypointrhs}[1]{\ensuremath{r^{\chi(#1)}}} 
\newcommand{\defbodypointdotrhs}[1]{\ensuremath{r^{\zeta(#1)}}} 

\newcommand{\ns}{\ensuremath{n_{s}}} 
\newcommand{\nq}{\ensuremath{n_{q}}} 
\newcommand{\nqx}{\ensuremath{{n}_{u}}} 
\newcommand{\nqy}{\ensuremath{{n}_{v}}}
\newcommand{\nf}{\ensuremath{n_{f}}} 
\newcommand{\nsd}{\ensuremath{n_{sd}}} 
\newcommand{\nneighbors}{\ensuremath{n_{p}}} 

\newcommand{\bodypointx}[1]{\ensuremath{\xi_{#1}}}
\newcommand{\bodypointy}[1]{\ensuremath{\eta_{#1}}} 
\newcommand{\flowpointx}[1]{\ensuremath{x_{#1}}}
\newcommand{\flowpointy}[1]{\ensuremath{y_{#1}}} 
\newcommand{\sdpointx}[1]{\ensuremath{x^{0}_{#1}}}
\newcommand{\sdpointy}[1]{\ensuremath{y^{0}_{#1}}}
\newcommand{\weights}[1]{\ensuremath{w_{#1}}} 

\newcommand{\bB}{\ensuremath{\B}}  
\newcommand{\bEo}{\ensuremath{\Eo}}

\newcommand{\bA}{\ensuremath{\A}}  
  
\newcommand{\bIq}{\ensuremath{\I_{q}}}
\newcommand{\bIs}{\ensuremath{\I_{s}}}  
\newcommand{\bC}{\ensuremath{\C}}  
\newcommand{\bCt}{\ensuremath{\Ct}}  
\newcommand{\bS}{\ensuremath{S}}
\newcommand{\bSt}{\ensuremath{S^{T}}}  
\newcommand{\lambdatrue}{\ensuremath{{\Lambda}}}

\newcommand{\Uc}{\ensuremath{{U}_{c}}}  
\newcommand{\Ucx}{\ensuremath{{F}_{u}}}  
\newcommand{\Ucy}{\ensuremath{{F}_{v}}}

\newcommand{\bstress}{\ensuremath{{f}}}
\newcommand{\bPnk}[2]{\ensuremath{{P}_{#1}^{(#2)}}}

\newcommand{\bodypointn}[1]{\ensuremath{\chi_{#1}}}
\newcommand{\noslipn}[1]{\ensuremath{\bm{u}_{r}^{#1}}}

\newcommand{\Sx}{\ensuremath{{S}_{x}}}
\newcommand{\Sy}{\ensuremath{{S}_{y}}} 
\newcommand{\Cx}{\ensuremath{{C}_{x}}}
\newcommand{\Cy}{\ensuremath{{C}_{y}}} 
\newcommand{\Cxiii}{\ensuremath{{C}_{x}^{(3)}}}

\newcommand{\droptol}{\ensuremath{\epsilon}}

\begin{document}

\begin{frontmatter}

\title{A strongly coupled immersed boundary method for fluid-structure interaction that mimics the efficiency of stationary body methods}


\author[rvt]{Nirmal~J.~Nair\corref{cor1}}% \fnref{fn1}}
\ead{njn2@illinois.edu}

\author[rvt]{Andres~Goza}%\fnref{fn2}}

\cortext[cor1]{Corresponding author}
%\fntext[fn1]{Graduate Research Assistant}
%\fntext[fn2]{Assistant Professor}

\address[rvt]{Department of Aerospace Engineering, University of Illinois at Urbana-Champaign, Urbana, IL 61801 USA}

\begin{abstract}

Strongly coupled immersed boundary (IB) methods solve the nonlinear fluid and structural equations of motion simultaneously for strongly enforcing the no-slip constraint on the body.
Handling this constraint requires solving several large dimensional systems that scale by the number of grid points in the flow domain even though the nonlinear constraints scale only by the small number of points used to represent the fluid-structure interface. 
These costly large scale operations for determining only a small number of unknowns at the interface creates a bottleneck to efficiently time-advancing strongly coupled IB methods. 
In this manuscript, we present a remedy for this bottleneck that is motivated by the efficient strategy employed in stationary-body IB methods while preserving the favorable stability properties of strongly coupled algorithms---we precompute a matrix that encapsulates the large dimensional system so that the prohibitive large scale operations need not be performed at every time step. This precomputation process yields a modified system of small-dimensional constraint equations that is solved at minimal computational cost while time advancing the equations. We also present a parallel implementation that scales favorably across multiple processors. The accuracy, computational efficiency and scalability of our approach are demonstrated on several two dimensional flow problems. Although the demonstration problems consist of a combination of rigid and torsionally mounted bodies, the formulation is derived in a more general setting involving an arbitrary number of rigid, torsionally mounted, and continuously deformable bodies.

\end{abstract}




\end{frontmatter}

%=======================================================================================

\section{Introduction}

Immersed boundary (IB) methods are numerical techniques for simulating the flow around bodies. In this framework, the bodies described by Lagrangian points are immersed into the fluid domain discretized by non-body-conforming Eulerian points. The interaction between the fluid and body is achieved via interpolation, which allows for the no-slip condition on the immersed body to be enforced by localized momentum forcing near the body. For flows past bodies that are stationary or undergoing prescribed kinematics, the interpolation operators relating the fluid and structure can be formulated to be independent of time. In this setting, the stresses on the immersed surface that enforce the no-slip constraint can be efficiently obtained via small-dimensional, time-constant linear systems with matrices that can be precomputed before advancing the equations in time \cite{mittal2005immersed,kim2006immersed,taira2007immersed}. However, in fully coupled fluid-structure interaction (FSI) problems, the unknown structural motion leads to a nonlinear algebraic constraint with time-varying operators that can no longer be efficiently precomputed      \cite{kim2019immersed,huang2019recent}.

There are a number of ways to handle this nonlinear constraint arising from the fluid-structure coupling. Weakly coupled IB methods treat the body forces or no-slip constraint explicitly in time. Although this approach removes the need to iterate on a nonlinear system of equations to advance the system in time \cite{kim2018weak,wang2020immersed}, the explicit treatment can impose severe time step restrictions if the structure undergoes  large deformations or if the structure-to-fluid mass ratio is low \cite{causin2005added,forster2007artificial,borazjani2008curvilinear}. 

By contrast, strongly coupled IB methods treat the body forces and no-slip constraints implicitly in time, allowing for stable simulations of FSI systems with modest time step sizes. 
The implicit treatment in these strongly coupled methods necessitates that the fluid and structural equations as well as the nonlinear interface constraint be solved simultaneously  via an iterative scheme \cite{tian2014fluid,de2016moving,degroote2009performance}.
These iterative approaches require, for each FSI iteration, the solution of a large system of equations involving not only the nonlinear constraint and but also the structural and flow equations that scale with the large number of points in the flow domain. These additional large linear solves, which are not present in the stationary-body setting, arise because of the small-dimensional, time-dependent no-slip condition that scales with the number of points on the fluid-structure interface. The small-dimensional nature of this FSI coupling offers a tantalizing question: can the additional expense, compared with stationary-body problems, of time-advancing fully coupled FSI systems be restricted to small-dimensional systems that scale with the number of points at the fluid-structure interface where the FSI coupling occurs?

Towards this aim, some IB methods have reformulated the fully coupled system of equations via block Gauss-Seidel \cite{wang2015strongly} or block-LU factorization \cite{goza2017strongly}, so that the iterations are restricted only to the variables existing on the fluid-structure interface. Yet, a key bottleneck to cost reductions in these reformulations is that there is inevitably a  large linear system---that scales with the large number of unknowns in the entire flow domain---that gets embedded within the small dimensional nonlinear FSI coupling equation. 
%Merely constructing the small-dimensional matrix is computationally expensive since it entails several computations involving the large embedded system. Furthermore, this small coupling matrix is dependent on the time-varying position of the immersed body, and therefore it must be constructed at least once per FSI iteration. The process of constructing the small-dimensional matrix at each FSI iteration therefore dominates the computational cost of these IB methods, only to eventually yield the small-dimensional body variables. We note that, for flows past stationary bodies, the small dimensional coupling matrix is not time dependent. This allows one to precompute the matrix once and for all in the beginning of the simulation and the above described computational bottleneck is not encountered \cite{kim2006immersed,colonius2008fast}.

A similar embedding of a large linear system within a small-dimensional matrix is also observed in some non-iterative IB methods \cite{lacis2016stable}. These methods utilize a semi-explicit treatment of the body forces or the no-slip constraint, with the benefit that the system may be advanced in time without iteration. Moreover, these approaches have been demonstrated to have favorable stability properties compared with weakly coupled methods, and are therefore often also referred to as strongly coupled methods. However, in the current work we refer to these methods as semi-strongly coupled because they do not strictly enforce the nonlinear algebraic constraint at a given time step, and often result in a reduction in the temporal accuracy of the solver to first order\footnote{We note that some semi-strongly coupled IB methods \cite{tschisgale2020immersed,yang2015non,xu2018novel} do not have an embedding of the large system due to their specific formulations. However, these methods have a reduced first order temporal accuracy due to the semi-explicit treatment of boundary constraints.}.

In this article, we focus on these strongly and semi-strongly coupled methods that contain an embedded large system within the small nonlinear FSI coupling equation because of their favorable stability properties and potential for computational efficiency. We note that the embedded large-dimensional solve provides a significant obstacle to any practical benefits associated with the nominally small-dimensional nature of the algebraic systems to be iterated on: merely constructing the small-dimensional matrix is computationally expensive since it entails several linear solves involving the large embedded system. Furthermore, this small coupling matrix is dependent on the time-varying position of the immersed body, and therefore it must be constructed at least once per time step (semi-strongly coupled methods) or once per FSI iteration (strongly-coupled methods). The process of constructing the small-dimensional matrix therefore dominates the computational cost of time-advancing these IB methods. We emphasize that this costly process is in contrast to that for flows past stationary bodies, where the small dimensional coupling matrix is not time dependent. This time independence allows one to precompute the coupling matrix once at the beginning of a simulation, allowing for the full system to be advanced without the bottleneck described above \cite{kim2006immersed,colonius2008fast}.

We present an efficient remedy for addressing the embedded large linear solve, towards realizing an iterative time advancement scheme that makes use of the small dimensional nature of the FSI coupling. The proposed approach preserves the favorable stability properties of these strongly and semi-strongly coupled schemes, while mimicking desirable features of the stationary body setting -- namely, precomputing a matrix that encapsulates the large linear system so that the several prohibitive large linear solves need not be performed at every time step. 
We also describe a parallel implementation of our FSI algorithm and demonstrate favorable strong scaling on a relatively large two-dimensional problem. Our formulation is developed for FSI problems involving an arbitrary number of rigid, torsionally mounted, and elastically deformable bodies, though for simplicity of presentation our results focus on a combination of rigid and torsionally mounted bodies.

The remainder of the paper is organized as follows. In Sec. \ref{background}, we give a background of the IB method of \citet{goza2017strongly}, which serves as the basis for the specific algorithm proposed in this article. We emphasize that the proposed approach for efficiently addressing the embedded large linear solve arising from many FSI systems has applicability beyond \citet{goza2017strongly}. To demonstrate this fact, we further describe in Sec. \ref{background} how the aforementioned bottleneck appears in a number of semi-strongly coupled and strongly coupled methods. The proposed efficient treatment of the FSI coupling is detailed in Sec. \ref{proposed}, and the strategies for parallel implementation on multiple processors are discussed in Sec. \ref{parallelimplementation}. We demonstrate the accuracy, computational efficiency and scalability of our approach on several two-dimensional (2D) flow problems in Sec. \ref{experiments}. Finally, conclusions are offered in Sec. \ref{conclusions}.

%=======================================================================================

\section{Background: strongly-coupled immersed boundary formulation}
\label{background}

In this section, we first review the strongly-coupled immersed boundary (IB) formulation by \citet{goza2017strongly}, discuss the source of the computational bottleneck encountered by this approach, and demonstrate the appearance of this bottleneck in other semi-strongly coupled and strongly coupled numerical methods. In the next section, we will discuss the remedy to this bottleneck. %However, we will emphasize in Sec. \ref{bottleneck} that this bottleneck is not limited to only this IB method and describe the class of IB methods affected by this issue. 

\subsection{Governing equations}

\begin{figure}
\centering
\includegraphics[scale=1]{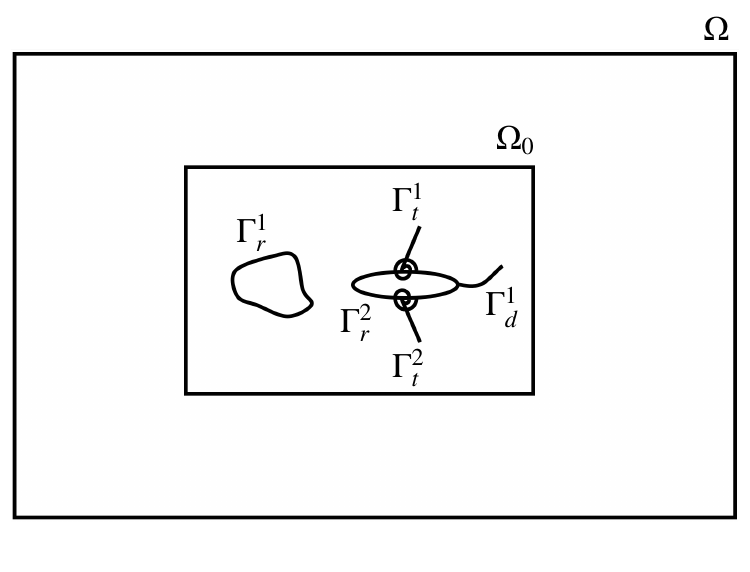}
\caption{Schematic of the computational domain consisting of the flow domain, $\fluiddomain$ and five immersed bodies, $\body=\{\rigidbody^1,\rigidbody^2,\torsbody^1,\torsbody^2,\defbody^1\}$. Rigid bodies include $\rigidbody^1$ and $\rigidbody^2$ while $\torsbody^1$ and $\torsbody^2$ are torsional bodies and $\defbody^1$ is a deformable body mounted on $\rigidbody^2$.  The sub-domain of our proposed approach described in Sec. \ref{proposed} that encompasses the range of motion of all bodies is denoted by $\subdomain$.}
\label{subdomain}
\end{figure}

We consider a fluid domain $\fluiddomain$ and a set of immersed bodies $\body$. We present a formulation for FSI problems involving a collection of $\nbr$ rigid bodies, $\rigbodyi$ for $i=1,\dots,\nbr$, along with $\nbt$ torsional bodies $\torsbodyi$ for $i=1,\dots,\nbt$ and $\nbd$ deformable bodies $\defbodyi$ for $i=1,\dots,\nbd$. The torsional bodies are assumed to be mounted on some subset of the rigid bodies, as shown in Fig \ref{subdomain}. The readers are referred to \cite{wang2015strongly} for details about bodies that are torsionally connected to other torsional bodies. Incorporating this extension would involve only superficial changes to the formulation. The dimensionless governing equations are written as
\begin{equation}
\frac{\partial \velocity}{\partial \timet} + \velocity \cdot \nabla \velocity = -\nabla \pressure + \frac{1}{\reynolds}  \nabla^2 \velocity + \int_{\body} \bm{\stress}(\bodypoint(\surface,\timet)) \delta(\bodypoint(\surface,\timet)-\flowpoint)d\surface
\label{ns}
\end{equation}
\begin{equation}
\nabla \cdot \velocity = 0 
\label{continuity}
\end{equation}
\begin{equation}
\inertia \frac{\partial^2 \deflectioni}{\partial \timet^2} + \damper \frac{\partial \deflectioni}{\partial \timet} +\stiffness \deflectioni = - \int_{\torsbodyi} (\torsbodypoint-\hinge) \times \stress(\torsbodypoint) d\torsbodypoint + \  \dynamics(\deflectioni) \quad \quad \text{for} \quad i=1,\ldots,\nbt %\forall i\in \NN: \torsbodyi \in \torsbody
\label{newtons}
\end{equation}
\begin{equation}
\frac{\densitys}{\densityf}\frac{\partial^2 \defbodypoint}{\partial \timet^2} = \frac{1}{\densityf \velocityscale^2} \nabla\cdot \cauchystress + \dynamicsdef(\defbodypoint) - \stress(\defbodypoint) \quad \quad \text{for} \quad i=1,\ldots,\nbd %\forall i\in \NN: \defbodyi \in \defbody
\label{beam}
\end{equation}
\begin{equation}
\int_{\fluiddomain} \velocity(\flowpoint) \delta(\flowpoint-\rigbodypoint)d\flowpoint = \noslipn{i}(\rigbodypoint) \quad \quad \text{for} \quad i=1,\ldots,\nbr
\label{bcr}
\end{equation}
\begin{equation}
\int_{\fluiddomain} \velocity(\flowpoint) \delta(\flowpoint-\torsbodypoint)d\flowpoint = \frac{\partial \deflectioni}{\partial t} \hat{\bm{e}}^i \times (\torsbodypoint-\hinge) \quad \quad \text{for} \quad i=1,\ldots,\nbt
\label{bct}
\end{equation}
\begin{equation}
\int_{\fluiddomain} \velocity(\flowpoint) \delta(\flowpoint-\defbodypoint)d\flowpoint = \frac{\partial \defbodypoint}{\partial \timet} \quad \quad \text{for} \quad i=1,\ldots,\nbd
\label{bcd}
\end{equation}
In the above, $\flowpoint$ denotes the Eulerian coordinate representing a position in space and $\bodypoint(\surface, \timet)$ denotes the Lagrangian coordinate attached to the bodies in the set $\body$, the surface of which is parametrized by the variable $\surface$. These variables, $\flowpoint$, $\bodypoint$ and $\surface$ were nondimensionalized by a characteristic length scale $\lengthscale$; velocity $\velocity$ was nondimensionalized by a characteristic velocity scale $\velocityscale$; time $\timet$ was nondimensionalized by $\lengthscale/\velocityscale$; pressure $\pressure$ and surface stress imposed on the fluid by the body $\stress$ were nondimensionalized by $\densityf \velocityscale^2$, where $\densityf$ is the fluid density. The Reynolds number in Eq. \eqref{ns} is defined as $\reynolds = \velocityscale \lengthscale/\visc$, where $\visc$ is the kinematic viscosity of the fluid. 

The equation of motion of the $i^{th}$ torsional body $\torsbodyi$ is given by Eq. \eqref{newtons} where $\deflectioni$ is the deflection angle of the body from it's undeformed configuration $\deflectionoi$ and $\torsbodypoint$ is the Lagrangian coordinate of $\torsbodyi$. Here, $\inertia$ denotes the moment of inertia of the torsionally connected body about a hinge location $\hinge$ nondimensionalized as $\inertia = \inertiadim/\densityf\lengthscale^4$, where $\inertiadim$ is the dimensional moment of inertia. Similarly, the torsional spring has a nondimensional stiffness $\stiffness = \stiffnessdim/\densityf \velocityscale^2\lengthscale^2$ and damping coefficient $\damper = \damperdim/\densityf\velocityscale\lengthscale^3$, where $\stiffnessdim$ and $\damperdim$ are the dimensional quantities, respectively. The first term on the right hand side of Eq. \eqref{newtons} represents the moment about $\hinge$ due to the surface stress imposed on the fluid by the body (thereby resulting in a negative sign). The second term $\dynamics$ represents moments due to body forces such as gravity, pseudo forces, \emph{etc.} %While only rigid and torsional bodies are considered here for simplicity, the IB algorithm of \cite{goza2017strongly} is agnostic to the type of bodies, and can be used without modification provided that the appropriate structural equations of motion are substituted for Eq. \eqref{newtons}.

The equation of motion of the $i^{th}$ deformable body $\defbodyi$ is given by Eq. \eqref{beam} where $\defbodypoint$ is the Lagrangian coordinate of $\defbodyi$. Here $\densitys$ is the density of the structure, $\cauchystress$ is the Cauchy stress tensor contributing to the internal restoring forces of the body and $\dynamicsdef$ denotes the body force per unit volume due to gravity, psuedo forces etc. See reference \cite{goza2017strongly} for a detailed description about these quantities.

The no-slip boundary constraints on the rigid, torsional and deformable bodies are given by Eq. \eqref{bcr}, \eqref{bct} and \eqref{bcd}, respectively. Here, $\noslipn{i}$ is the (possibly zero) prescribed velocity on the rigid body $\rigbodyi$, and $\hat{\bm{e}}^i$ is a unit vector denoting the direction of the angular velocity of the torsional body $\torsbodyi$. These no-slip constraints are used to solve for the surface stress term $\stress(\bodypoint)$ that enforces the boundary condition that must hold on the respective bodies.

%=======================================================================================

\subsection{Fully discretized equations}

Following \cite{goza2017strongly}, Eq. \eqref{ns} is spatially discretized using the standard second-order finite difference operators and rewritten in a streamfunction-vorticity formulation. A finite element procedure described in  \cite{goza2017strongly} is used to spatially discretize Eq. \eqref{beam}. For time-discretization, the flow equations \eqref{ns} utilize an Adams-Bashforth scheme for the nonlinear term and a Crank-Nicolson method for the diffusive term. The structural equations of motion \eqref{newtons} and \eqref{beam} are discretized using an implicit Newmark scheme. The boundary conditions \eqref{bcr}--\eqref{bcd} and the surface stress term in Eq. \eqref{ns} are evaluated implicitly at the current time step to enable stability of the method for bodies with a wide range of mass ratio and undergoing large body displacements. The fully discretized equations are given below,
\begin{equation}
\Ct\A\C \streamfnc_{n+1} + \Ct \Etn{n+1}\stressn{n+1} = \fluidrhs
\label{ns4}
\end{equation}
\begin{equation}
\frac{4}{\dt^2}\inertia\deflectioni_{n+1} + \frac{2}{\dt}\damper \deflectioni_{n+1} + \stiffness \deflectioni_{n+1} - \Qi{}\Ri{,n+1}{}{\stresst{,n+1}}\ds = r^{\deflectiondot,i}_{n}  \quad \quad \text{for} \quad i=1,\ldots,\nbt
\end{equation}
\begin{equation}
\frac{2}{\dt}\deflectioni_{n+1} - \deflectiondoti_{n+1} = r_n^{\deflection,i} \quad \quad \text{for} \quad i=1,\ldots,\nbt
\end{equation}
\begin{equation}
\frac{4}{\dt^2}\Mdi\defbodypointi{,n+1} + \Rdi(\defbodypointi{,n+1}) - \Qdi\Wdi{,n+1}{}\stressd{,n+1} = r^{\zeta,i}_{n}  \quad \quad \text{for} \quad i=1,\ldots,\nbd
\label{beam4}
\end{equation}
\begin{equation}
\frac{2}{\dt}\defbodypointi{,n+1} - \defbodypointdoti{,n+1} = r_n^{\bodypointn{},i} \quad \quad \text{for} \quad i=1,\ldots,\nbd
\end{equation}
\begin{equation}
\En{r,n+1}^{i} \C \streamfnc_{n+1} = u_{r,n+1}^i  \quad \quad \text{for} \quad i=1,\ldots,\nbr
\end{equation}
\begin{equation}
\En{t,n+1}^{i} \C \streamfnc_{n+1} - \Ri{,n+1}{T}\Qi{T} \deflectiondoti_{n+1} = 0  \quad \quad \text{for} \quad i=1,\ldots,\nbt
\end{equation}
\begin{equation}
\En{d,n+1}^{i} \C \streamfnc_{n+1} - \defbodypointdoti{,n+1} = 0  \quad \quad \text{for} \quad i=1,\ldots,\nbd
\label{bc4}
\end{equation}
Here, the subscript $n$ denotes the time step; the discrete streamfunction and surface stresses imposed on all bodies by fluid are denoted by $\streamfnc$ and $\stressn{}$, respectively; the stresses on the individual torsional and deformable bodies are denoted by $\stresst{}, \stressd{}\in \stressn{}$, respectively (there is also a set of surface stresses associated with the rigid bodies, $f_r^i\in\stressn{}$). We also define $\deflectiondoti = \dot{\deflection}^i$ and  $\defbodypointdoti{} =\dot{\chi}^i_d$. The curl operator is given by $\C$; $\A = \frac{1}{\dt}\I - \frac{1}{2}\LL$ where $\dt$ is the time step size, $\I$ is the identity and $\LL$ is the vector Laplacian operator. The discretization of the operators in the left hand side of Eq. \eqref{bcr}--\eqref{bcd}, $\En{r}^i$, $\En{t}^i$, and $\En{d}^i$, are the IB interpolation operators that interpolate the fluid velocity onto the rigid, torsional and deformable bodies, respectively and $\En{}$ is simply the block-row aggregation of each of $\En{r}^i$, $\En{t}^i$, and $\En{d}^i$. On the other hand, $\Etn{}$ represents the regularization operator involving the delta function in Eq. \eqref{ns} which regularizes surface stress from each of the bodies onto the flow field. See reference \cite{taira2007immersed} for more details about the standard finite volume discretizations used to represent fluid operators (e.g., $C$, $L$) as well as more information on the IB interpolation and regularization operators.

The operator $\Qi{}\Ri{,n+1}{}$ denotes the discretization of the term involving the surface stress in Eq. \eqref{newtons} and $\ds$ is the size of discretization of the body while $\Mdi$, $\Rdi$ and $\Qdi{}\Wdi{,n+1}{}$ are the finite element operators corresponding to the first, second and fourth terms of Eq. \eqref{beam}. See Appendix A for the more details of these operators. The expressions of the right hand side terms $\fluidrhs$, $r^{\deflectiondot,i}_{n}$, $r_n^{\deflection,i}$, $r^{\zeta,i}_{n}$ and $r_n^{\bodypointn{},i}$, which are the known right-hand side quantities that arise from the explicit temporal treatment and boundary conditions, are also provided in Appendix A.

%=======================================================================================

\subsection{Algorithm for strong fluid-structure coupling and associated computational bottleneck}
\label{bottleneck}

The implicit treatment of body variables and the no-slip constraint in strongly-coupled IB methods necessitates an iterative method to solve the above system of equations \eqref{ns4}--\eqref{bc4}.
However, a straightforward implementation of iterating all the equations until convergence will incur significant expense since the flow equations scale by the large number of flow points. An observation of \citet{goza2017strongly} was that Eq. \eqref{ns4}--\eqref{bc4} can be subjected to a block-LU decomposition before applying an iterative scheme so that the iterations are only restricted to the evaluation of small-dimensional systems that scale only with the number of points on the immersed surface. The full derivation of this procedure is provided in Appendix A for self-containment. The final system of LU-factored equations is 

%To enable strong fluid-structure coupling, an iterative procedure to determine the surface stress and position of the body is introduced into the fully discretized equations. Finally, the system of equations are rewritten in a fractional step form by performing a block-LU factorization of the original linear system. For self-containment, the full derivation of this strongly-coupled projection based immersed boundary approach is provided in Appendix A. The final system of equations is given below,
%
\begin{equation}
\streamfnc^* = (\Ct \A \C)^{-1} \fluidrhs
\label{predictor}
\end{equation}
\begin{equation}
\begin{split}
\left( \Enk{n+1}{k} \C (\Ct\A\C)^{-1}\Ct\Etnk{n+1}{k} + \frac{2\ds}{\dt} \Stnk{n+1}{k} \J^{-1} \Snk{n+1}{k} + \frac{2}{\dt} \hat{I}_{d}^T {\Jd}^{-1} \Sdnk{n+1}{k} \right) \stressnk{n+1}{k+1} = \\
\Enk{n+1}{k}\C\streamfnc^* - \bcrhs{k} + \Stnk{n+1}{k} \left( \deflectionrhs{k} - \frac{2}{\dt}\J^{-1}\deflectiondotrhs{k} \right) +  \hat{I}_{d}^T \left( \defbodypointrhs{k} - \frac{2}{\dt}{\Jd}^{-1}\defbodypointdotrhs{k} \right)
\label{stress}
\end{split}
\end{equation}
\begin{equation}
\deflectionincrease = \J^{-1} \left( \deflectiondotrhs{k} + \ds\Snk{n+1}{k} \stressnk{n+1}{k+1} \right)
\label{dtheta}
\end{equation}
\begin{equation}
\defbodypointincrease = {\Jd}^{-1} \left( \defbodypointdotrhs{k} + \Sdnk{n+1}{k} \stressnk{n+1}{k+1} \right)
\label{dchi}
\end{equation}
\begin{equation}
\streamfnc_{n+1} = \streamfnc^{*} - (\Ct\A\C)^{-1} \Ct\Etn{n+1}\stressn{n+1}
\label{corrector}
\end{equation}
Here, the superscript $(k)$ denotes the FSI iteration. $\Snk{n+1}{k}$, $\Sdnk{n+1}{k}$, $\J$, $\Jd$ and $\hat{I}_d$ are the aggregated block-diagonal matrices containing the individual structural operators with the superscript $i$ in \eqref{ns4}-\eqref{bc4}. The expressions of these block-diagonal operators as well as the right-hand side terms $\fluidrhs$, $\bcrhs{k}$, $\deflectionrhs{k}$, $\deflectiondotrhs{k}$, $\defbodypointrhs{k}$ are provided in Appendix A.
%The operator $\Sn{}$ denotes the discretization of the term involving surface stress in Eq. \eqref{newtons}; $\J$ is a square diagonal operator of size of the number of torsional bodies with diagonal elements $\J_{(i,i)} = \frac{4}{\dt^2}\inertia + \frac{2}{\dt}\damper + \stiffness$ and $\ds$ is the size of discretization of the body

Now, the entire method can be efficiently divided into three steps. First, a trial streamfunction $\streamfnc^*$ is predicted without accounting for the body forces in Eq. \eqref{predictor}. Next,  the FSI coupling Eq. \eqref{stress}--\eqref{dchi} are solved iteratively at the next time step $n+1$ for the surface stress $\stressnk{n+1}{k+1}$ and body configuration
$\deflectionnk{n+1}{k+1}=\deflectionnk{n+1}{k} + \deflectionincrease$, 
$\defbodypointnk{n+1}{k+1}=\defbodypointnk{n+1}{k} + \defbodypointincrease$.
Within each FSI iteration, the linear system in Eq. \eqref{stress} is solved using an iterative method such as GMRES. We note that there is a distinction between the FSI iterations associated with Eq. \eqref{stress}--\eqref{dchi} and the GMRES iterations used to solve Eq. \eqref{stress} within each FSI iteration. We will differentiate between these two types of iterations as needed for clarity of context. Finally, the streamfunction at the current time step, $\streamfnc_{n+1}$, is obtained by correcting $\streamfnc^*$ using the updated surface stress in Eq. \eqref{corrector}.

We note that the trial and corrected streamfunctions and therefore Eq. \eqref{predictor} and \eqref{corrector} scale by the large number of flow points. However, Eq. \eqref{predictor} and \eqref{corrector} do not depend on the FSI iterate, $k$, and therefore, are solved only once at the beginning and end of the time-step, respectively. These steps thus incur the same cost as compared to the non-FSI, stationary body case, which is a lower bound for the computational expense one can expect to obtain for fully coupled FSI simulations. In contrast, Eq. \eqref{stress}--\eqref{dchi} are solved for multiple FSI iterates $k$ within a single time step. Since the system of iterated equations \eqref{stress}--\eqref{dchi} are small dimensional scaling by the number body points, nominally a significant amount of computational savings can be expected as compared to a straightforward implementation of iterating over all the equations.

%=======================================================================================

%\subsection{Outline: Computational bottleneck in strongly-coupled FSI methods}
%\label{bottleneck}

These savings are realized due to the block-LU decomposition of Eq. \eqref{ns4}--\eqref{bc4}. However, an undesirable consequence of this decomposition procedure is that a large linear system in the form on $(C^TAC)^{-1}$ that scales with the number of points in the flow domain, $\ns$, gets embedded within the small dimensional matrix in Eq. \eqref{stress}, $\Enk{n+1}{k} \C (\Ct\A\C)^{-1}\Ct\Etnk{n+1}{k}$.
%, has an embedded large linear system in the form on $(C^TAC)^{-1}$ that scales with the number of points in the flow domain, $\ns$. 
Merely constructing the small-dimensional matrix $\Enk{n+1}{k} \C (\Ct\A\C)^{-1}\Ct\Etnk{n+1}{k}$ is computationally expensive since it requires several large computations involving $(C^TAC)^{-1}$. Furthermore, this small matrix depends on the time-varying position of the immersed body, and therefore changes at least once per FSI iteration within each time step.

The full construction of $\Enk{n+1}{k} \C (\Ct\A\C)^{-1}\Ct\Etnk{n+1}{k}$ may be  circumvented by a matrix-free implementation of GMRES. However, even in the matrix-free implementation, these $(\Ct\A\C)^{-1}$ operations are performed once in every GMRES iteration within every FSI iteration. For example, if the algorithm requires 3 FSI iterations per time step and on average, each FSI iteration requires 5 GMRES iterations, then a total of $3 \times 5=15$ operations of $(\Ct \A \C)^{-1}$ are performed just in a single time step. Therefore, even though the underlying linear system in \eqref{stress} is small dimensional, multiple solves of the large embedded system is inevitable.

%We note that although the bottleneck of embedding of a large linear system within a small matrix arises from the block LU decomposition, the root cause of the bottleneck is the implicit treatment of body forces, positions and no-slip constraint.

The root cause for this bottleneck is the need to solve the system of equations \eqref{ns4}--\eqref{bc4} simultaneously arising from the implicit treatment of body forces, positions and no-slip constraint in strongly-coupled methods. This implicit treatment necessitates the computation of the surface stress such that it enforces the no-slip constraint at the current time step.
In the IB method of \citet{goza2017strongly}, this implicit treatment is manifested in the first term of Eq. \eqref{stress} as
\begin{equation}
\overset{\text{Contribution to no-slip velocity}}{\overbrace{\Enk{n+1}{k} \C \quad \overset{\text{Globally affected flow-field}}{\overbrace{(\Ct\A\C)^{-1} \quad \overset{\text{Local fluid source}}{\overbrace{\Ct\Etnk{n+1}{k} \stressnk{n+1}{k+1}} } } } } }
\end{equation}
A source term in the form of surface stress $\stressnk{n+1}{k+1}$ is converted into a local fluid source in the vicinity of the Lagrangian body points via the action of $\Ct\Etnk{n+1}{k}$. Then the elliptic Poisson-like operator $(\Ct \A \C)^{-1}$ containing the viscous contribution globally modifies the flow-field. Finally, the no-slip velocity on the body enforced by the surface stress is obtained via interpolation of the globally affected flow-field via $\Enk{n+1}{k} \C$. %In summary, irrespective of the small size of surface stress, the intermediate expensive Poisson-like solve in Eq. \eqref{stress}, emerging due to the implicit treatment of body variables and no-slip constraint, creates a severe computational bottleneck. 

We emphasize that the above-mentioned bottleneck of solving a large-dimensional system for the small dimensional body variables is not limited to the IB method of \citet{goza2017strongly}. Broadly speaking, fully implicit, strongly coupled IB methods require iterations to arrive at a solution that satisfies the flow and structural equations of motion as well as the nonlinear no-slip constraint. Many iterative approaches require iterating on all flow (velocity, pressure) and structural (displacement, forces) variables, the former of which requires the solution of large-dimensional systems that scale with the number of flow points \citep{tian2014fluid,degroote2009performance,de2016moving}. Other approaches more similar to that of \citet{goza2017strongly} are able to reformulate the discrete equations, through either a block Gauss-Seidel approach \cite{wang2015strongly} or a block-LU factorization \cite{lacis2016stable}, so that any required iterations are restricted to nominally small dimensional systems in analogy with \eqref{stress}--\eqref{dchi}. However, similar to the algorithm of \citet{goza2017strongly}, these small dimensional systems that scale with the number of body points at the fluid-structure interface have embedded large linear systems that scale with the number of points in the flow domain. 

%A remedy to the bottleneck is to revoke the implicit treatment of the body variables and no-slip constraint. However, this remedy will also result in a weakly-coupled IB method which can be undesirable due to the unfavourable stability properties in the presence of large deformations or low mass ratios.

In the next section, we propose an efficient algorithm that addresses the above-mentioned bottleneck of all strongly coupled and some semi-strongly coupled IB methods. Our proposed approach leverages the block-LU factored form of the equations \eqref{predictor}--\eqref{corrector}, so that the FSI iterations are restricted to small dimensional systems. We provide a strategy to precompute the matrix that encapsulates the large linear system on a sub-domain that envelops the full range of structural motion. The matrix is then updated to accurately enforce the no-slip constraint via interpolation onto the portion of the sub-domain for the location of the current structures. The precomputation procedure avoids additional large linear solves (compared to the non-FSI, stationary body case) while marching in time, while the interpolation procedure allows for accurate treatment of arbitrarily large structural motions. 

%The algorithm we provide below addresses the additional linear solves (compared with the rigid, stationary body case) that are required to time march all strongly coupled and most semi-strongly coupled IB methods. Our proposed approach leverages the ability to reformulate the system of equations, without approximation error, so that the FSI iterations are restricted to small dimensional systems. We provide a strategy for pre-computing a ...  
%=======================================================================================

\section{Proposed approach for treating arbitrarily moving bodies as efficiently as stationary bodies}
\label{proposed}

%In this section, we propose a novel and efficient sub-domain approach to tackle the above-mentioned computational bottleneck of performing multiple $(\Ct \A \C)^{-1}$ operations in Eq. \eqref{stress}. 

Our proposed idea is motivated from the observation that for \emph{stationary} bodies, the operator $\Enk{n+1}{k}$ and therefore, the small-dimensional operator $\Enk{n+1}{k} \C (\Ct\A\C)^{-1}\Ct\Etnk{n+1}{k}$ do not vary in time. This allows one to compute $\Enk{n+1}{k} \C (\Ct\A\C)^{-1}\Ct\Etnk{n+1}{k}$ once and for all, thereby circumventing the need to compute the computationally expensive $(\Ct\A\C)^{-1}$ at every GMRES iteration within each time step. Similarly, to avoid computing $(\Ct\A\C)^{-1}$ multiple times in Eq. \eqref{stress} for \emph{non-stationary} bodies, we propose the following approximation for $\Enk{n+1}{k}$,
\begin{equation}
\Enk{n+1}{k} \approx \Pnk{n+1}{k} \Eo
\label{sdapprox}
\end{equation}
where $\Eo$ is an IB interpolation operator similar to $\Enk{n+1}{k}$, but defined on a \emph{sub-domain} defined as a fixed set of $\nsd$ Eulerian points in the flow domain, $\subdomain\subset\fluiddomain$ as shown in Fig. \ref{subdomain}. These sub-domain points are selected \emph{a priori}, independent of the time-instantaneous body locations, and therefore $\Eo$ is time-invariant. The actual IB interpolation operator $\Enk{n+1}{k}$ defined on the moving Lagrangian body points is then recovered by the application of an interpolation operator  $\Pnk{n+1}{k}$ (not the same as the IB interpolation operator $\En{}$) on $\Eo$. This operator $\Pnk{n+1}{k}$ is time varying but may be evaluated sparsely and cheaply, as it only involves a small number of nonzero interpolation weights near the various structural interfaces (contained within the sub-domain). More details about $\Eo$ and  $\Pnk{n+1}{k}$ are discussed in Sec. \ref{interpolationEo}. For now, the previously expensive operation of Eq. \eqref{stress} can be rewritten as,
\begin{equation}
\Enk{n+1}{k} \C (\Ct\A\C)^{-1}\Ct\Etnk{n+1}{k} \approx \Pnk{n+1}{k} \left( \Eo \C (\Ct\A\C)^{-1}\Ct \Eot \right)\Ptnk{n+1}{k} = \Pnk{n+1}{k} \B \Ptnk{n+1}{k}
\label{operatorapprox}
\end{equation}
where $\B = \Eo \C (\Ct\A\C)^{-1}\Ct \Eot$. The above reformulation facilitates the following:
\begin{itemize}
\item[a.] Since $\B$ is time-invariant and scales by a size smaller than the flow points, $\nsd < \ns$ (often $\nsd \ll \ns$ depending on the range of the bodies' motions), we can compute and store $\B$ once and for all, thereby circumventing multiple $(\Ct\A\C)^{-1}$ operations.
\item[b.] Additionally, since $\Pnk{n+1}{k}$ is sparse, evaluation of $\Pnk{n+1}{k} \B \Ptnk{n+1}{k}$ is performed at minimal computational cost that scales only with a small multiple of the number of body interface points.
\end{itemize}

%=======================================================================================
\subsection{Sub-domain IB interpolation operator, \texorpdfstring{$\Eo$}{}, and sparse interpolation operator \texorpdfstring{$\Pn{}$}{}}
%\subsection{Overview of the sub-domain IB interpolation operator, $\Eo$, and approximating the true interpolation operator $\En{}$}
\label{interpolationEo}

The IB interpolation operator is constructed from the regularized discrete delta-function \cite{mittal2005immersed,peskin2002immersed}. If we denote the discrete delta function as $\discretedelta(\cdot)$, then the interpolation operator for interpolating the velocity from a Eulerian flow point at $\flowpoint=(\flowpointx{j},\flowpointy{j})$ to a Lagrangian body point $\bodypoint = (\bodypointx{i},\bodypointy{i})$ is given (to within a scaling factor \cite{taira2007immersed}) by,% For instance, the delta-function term involving the surface stress is discretized (to within a scaling factor) as,
\begin{equation}
\En{ij} \equiv \discretedelta(\flowpointx{j} - \bodypointx{i}) \ \discretedelta(\flowpointy{j} - \bodypointy{i})
\label{ibinterp}
\end{equation}
Note that the time subscript $n+1$ and iteration superscript $k$ are dropped for neatness. In our proposed approach, however, we first define a sub-domain $\subdomain \subset \fluiddomain$ as shown in Fig. \ref{subdomain}  and define an associated set of points $\sdpoint \in \subdomain$. The procedure for selecting the sub-domain is provided in Sec. \ref{choosesd1}. In contrast to Eq. \eqref{ibinterp}, the sub-domain interpolation operator is now defined between the Eulerian flow point $\flowpoint=(\flowpointx{j},\flowpointy{j})$ and Eulerian flow sub-domain point $\sdpoint=(\sdpointx{i},\sdpointy{i})$ as,
\begin{equation}
\Eoij{ij} \equiv \discretedelta(\flowpointx{j} - \sdpointx{i}) \ \discretedelta(\flowpointy{j} - \sdpointy{i})
\label{ibsdinterp}
\end{equation}
This interpolation operator associated with the sub-domain $\subdomain$, \eqref{ibsdinterp}, may be precomputed at the fixed set of sub-domain points. The desired IB interpolation operator, $\En{}$, associated with the time varying body locations is then approximated through interpolation of the precomputed sub-domain interpolation operator, $\Eo$ via
\begin{equation}
\En{ij} \equiv \discretedelta(\flowpointx{j} - \bodypointx{i}) \ \discretedelta(\flowpointy{j} - \bodypointy{i}) \approx \sum_{k=1}^{\nsd} \weights{ik} \ \discretedelta(\flowpointx{j} - \sdpointx{k}) \ \discretedelta(\flowpointy{j} - \sdpointy{k}) \equiv \sum_{k=1}^{\nsd} \Pn{ik} \ \Eoij{kj}
\label{sdapprox2}
\end{equation}
where $\weights{ik}$ are the weights of interpolation which are stored in the operator $\Pn{}$. 

The expression \eqref{sdapprox2} is meant to be illustrative of the interpolation process. In practice, it is wasteful to utilize the entire sub-domain $\subdomain$ to construct the interpolation weights. Instead, we perform local interpolation using only a small  number of $\nneighbors \ll \nsd$ nearest neighboring sub-domain points  to the body-point. In particular, for approximating $\En{ij}$ at the $i^{th}$ body point $(\bodypointx{i},\bodypointy{i})$, we identify $\nneighbors$ nearest neighboring sub-domain points $(\sdpointx{i_k},\sdpointy{i_k})$ where $i_k \in \{1,\ldots,\nsd\}$ for $k=1,\ldots,\nneighbors$. In other words, $(\sdpointx{i_k},\sdpointy{i_k})$ represents the $k^{th}$ nearest neighbor point on the sub-domain associated with the $i^{th}$ body point $(\bodypointx{i},\bodypointy{i})$. Accordingly, the interpolation in Eq. \eqref{sdapprox2} can be locally performed as
\begin{equation}
\En{ij} \equiv \discretedelta(\flowpointx{j} - \bodypointx{i}) \ \discretedelta(\flowpointy{j} - \bodypointy{i}) \approx \sum_{k=1}^{\nneighbors} \weights{ii_k} \ \discretedelta(\flowpointx{j} - \sdpointx{i_k}) \ \discretedelta(\flowpointy{j} - \sdpointy{i_k}) \equiv \sum_{k=1}^{\nneighbors} \Pn{ii_k} \ \Eoij{i_kj}
\label{sdapprox3}
\end{equation}
where now only $\weights{ii_k}$ needs to be stored for the body index $i$ and $\weights{il}=0 \ \ \forall \ l\in \{1,\ldots,\nsd\}, l \neq i_k \ \ \text{for} \ \ k=1,\ldots,\nneighbors$. 
%where $(\sdpointx{i_k},\sdpointy{i_k})$ represent the $k^{th}$ nearest neighbor point on the sub-domain associated with the $i^{th}$ body point $(\bodypointx{i},\bodypointy{i})$.

In this way, we may consider only the number of nearest neighbors, $\nneighbors$, in constructing and applying $\Pn{}$, rather than the total number of points in the sub-domain, $\nsd$. This formulation allows for $\Pn{}$, which is time dependent, to be efficiently constructed and applied via sparse operations.  The procedure for identifying the $\nneighbors$ sub-domain points nearest to a body point is provided in Sec. \ref{choosesd}. 

%=======================================================================================

\subsubsection{Procedure for selecting a sub-domain}
\label{choosesd1}

First, a rectangular sub-domain as shown in Fig. \ref{subdomain} is considered for simplicity. Next, the sub-domain boundaries are chosen such that all the bodies are guaranteed \emph{a priori} to stay within the sub-domain at all time instants. This can be achieved by examining the physical displacement limits of the body and total simulation time. A physical intuition of the problem can also help in choosing a more compact sub-domain. Since choosing the sub-domain is problem dependent, it will be discussed in more detail for specific problems in Sec. \ref{experiments}. Next, the grid spacing between the sub-domain points is set to be equal to the flow grid spacing. This choice was observed to provide accurate results for the set of problems considered in Sec. \ref{experiments}. Furthermore, in the staggered grid configuration, the sub-domain points are chosen to coincide with the vorticity points on cell vertices so that the sub-domain points are equidistant from the $x-$ and $y-$ velocity points located on the cell edges.

%=======================================================================================

\subsubsection{Choice of interpolation method}

A variety of interpolation functions such as Lagrange interpolation functions, delta functions, polynomial functions \emph{etc.}, can be used for performing interpolation and constructing $\Pn{}$ in Eq. \eqref{sdapprox3}. Since the use of delta functions for constructing $\En{}$ is well known and studied in the IB framework, we use delta functions for constructing $\Pn{}$ as well. We will denote these delta functions as $\discretedeltap(\cdot)$ and emphasize that the discrete delta functions $\discretedeltap(\cdot)$ used in $\Pn{}$ may be different from $\discretedelta(\cdot)$ used for constructing $\En{}$. While the choice of $\discretedelta(\cdot)$ is governed by the need to regularize and remove unphysical oscillations in surface stress \cite{goza2016accurate}, $\discretedeltap(\cdot)$ is chosen to strike a balance between the sparsity of $\Pn{}$ and accuracy of interpolation. 

In this work, we use a two-point hat function \cite{yang2009smoothing} given by
\begin{equation}
\discretedeltap(r) = \left\{\begin{matrix}
1-\frac{|r|}{\Delta r}, \quad |r|<\Delta r\\ 
0, \quad |r| > \Delta r
\end{matrix}\right.
\end{equation}
where $\Delta r$ is the flow sub-domain grid spacing in the $r$-direction. We choose this delta function because it has a support of only one cell and yet it is $\mathcal{O} (({\dsd})^2)$ accurate where $\dsd$ is the sub-domain grid spacing. A single cell support implies that for two dimensional IB method, only $\nneighbors=4$ input points are required for interpolation, thereby, enabling an extremely sparse construction of $\Pn{}$ with only $\nneighbors=4$ non-zeros per row. Furthermore, we note that the second order interpolation method does not affect the original first order spatial accuracy \cite{colonius2008fast} of projection based immersed boundary methods. Now, the weights of interpolation in Eq. \eqref{sdapprox3} are given by,
\begin{equation}
\weights{ii_k} = \discretedeltap(\bodypointx{i} - \sdpointx{i_k}) \ \discretedeltap(\bodypointy{i} - \sdpointy{i_k})
\label{weights}
\end{equation}

%=======================================================================================

\subsubsection{Choice of sub-domain points for local interpolation}
\label{choosesd}

\begin{figure}
\centering
\includegraphics[scale=1]{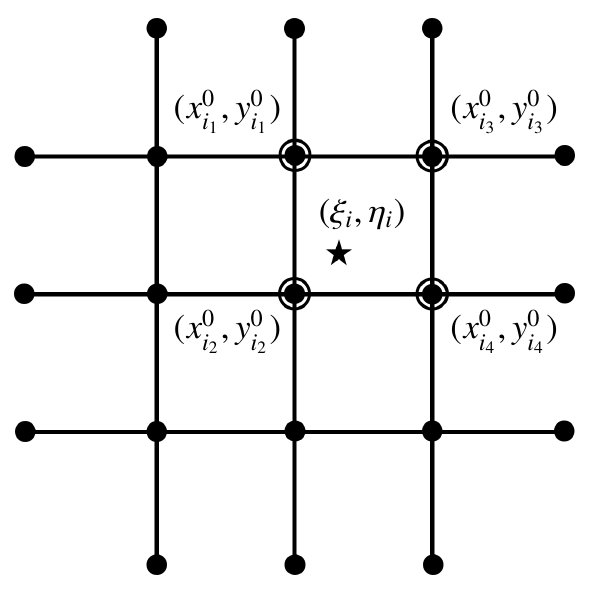}
\caption{Schematic for choosing $\nneighbors=4$ nearest neighboring sub-domain points $(\sdpointx{i_k},\sdpointy{i_k})$ for the body point $(\bodypointx{i},\bodypointy{i})$.}
\label{interppts}
\end{figure}

For local interpolation, $\nneighbors=4$ nearest neighboring sub-domain points that form a tensor grid are chosen. For instance, consider the sub-domain points in a two-dimensional space denoted by `$\bullet$' as shown in Fig \ref{interppts}. For approximating the operator $\En{}$ at the body point $(\bodypointx{i},\bodypointy{i})$ denoted by `$\star$', the four nearest neighboring points $(\sdpointx{i_k},\sdpointy{i_k})$ that form a tensor grid denoted by `$\circ$' are chosen.

%=======================================================================================

\subsection{Approximating \texorpdfstring{$\B$}{} as a sparse operator}
\label{sparsity}

The proposed sub-domain approach requires precomputing and storing the operator $\B$. However, we note that $\B$ is a dense matrix and therefore, storing $\B$ can become computationally prohibitive for problems with large sub-domain and fine grid discretization. To circumvent this computational storage issue, we approximate the dense $\B$ operator as sparse. In Sec. \ref{sparsity2}, we provide justification that $\B$ can be indeed constructed sparsely up to a drop tolerance. Then a drop tolerance filtering technique similar to that employed in incomplete LU decomposition \cite{jovanovic1970drop} to construct $\B$ sparsely is provided in Sec. \ref{filter}.

%=======================================================================================
\subsubsection{Analysis of sparsity of \texorpdfstring{$\B$}{}}
\label{sparsity2}

For clarity, we specify the dimensions of the previously defined operators as $\bEo \in \RR^{2\nsd \times \nq}$, $\bB \in \RR^{2\nsd \times 2\nsd}$, $\bC \in \RR^{\nq \times \ns}$ and $\bA \in \RR^{\nq \times \nq}$, respectively, where $\nsd$, $\nq$ and $\ns$ are the number of sub-domain grid points, sum of velocity grid points in $x$ and $y$ coordinate directions ($\nqx+\nqy$), and vorticity grid points, respectively. 

%It is noted that $\bEo=blockdiag(\bEox,\bEoy)$ where $\bEox \in \RR^{\nsd \times \nqx}$ and $\bEoy \in \RR^{\nsd \times \nqy}$ are the IB interpolation matrices that interpolate $x$ and $y$ velocity fields onto the sub-domain respectively. Furthermore, for the sake of analysis, we make two simplifications: (a) instead of a staggered grid, we consider a non-staggered grid where all flow quantities lie on the cell vertices, 
%thereby $\nqx=\nqy=\ns$, 
%and (b) we consider \emph{exact} delta functions with support of zero cells for the construction of $\bEo$, implying that $elements(\bEo)\in \{0,1\}$ since the sub-domain points originally coinciding with vorticity points, now also exactly coincide with the velocity grid points.

Firstly, we will focus on the interior term $\bC(\bCt \bA \bC)^{-1}\bCt$ of $\bB = \bEo\bC(\bCt \bA \bC)^{-1}\bCt\bEo^T$. Since $\bA = \bIq + \factor \bC\bCt$, where $\factor = \frac{\dt}{2\reynolds\dx^2}$, $\bIq \in \RR^{\nq \times \nq}$ is the identity and $\bC\bCt \in \RR^{\nq \times \nq}$ is the 2D vector Laplacian matrix, $\bCt \bA \bC$ can be rewritten as,
\begin{equation}
\bCt \bA \bC = \bCt\bC (\bIs + \factor \bCt\bC)
\end{equation}
where $\bIs \in \RR^{\ns \times \ns}$ is the identity and $\bCt\bC \in \RR^{\ns \times \ns}$ is the standard 2D scalar Laplacian. $\bCt\bC$ can be diagonalized as $\bCt \bC = \bS\lambdatrue\bSt$, where the eigenvectors $\bS \in \RR^{\ns\times\ns}$ are the discrete sine transforms and $\lambdatrue \in \RR^{\ns \times \ns}$ contains the eigenvalues. Accordingly, we can define the singular value decomposition, $\bC = \Uc \lambdatrue^{1/2} \bSt$ where $\Uc \in \RR^{\nq \times \ns}$ is the left singular vector. On substituting these decompositions, we get for the interior term,
\begin{equation}
\bC(\bCt \bA \bC)^{-1}\bCt = \Uc (\bIs + \alpha\lambdatrue)^{-1} \Uc^T
\label{sparseapprox1}
\end{equation}
For a conservative choice of grid Reynolds number $\reynolds \dx = 1$ and time discretization $\dt = \dx/4$ resulting in $\factor=0.125$, $\bIs + \alpha\lambdatrue$ has a small condition number of $2$. Therefore, we note that $\bIs + \alpha\lambdatrue$ is nearly a constant diagonal matrix (less conservative grid Reynolds numbers would only act to improve this approximation). Thus, $\Uc \Uc^T$ will have nearly the same sparsity structure as that of $\Uc (\bIs + \alpha\lambdatrue)^{-1} \Uc^T$. We therefore demonstrate below that $\Uc\Uc^T$ is well approximated as a sparse matrix, and use this to argue that the latter matrix $\Uc (\bIs + \alpha\lambdatrue)^{-1} \Uc^T$ will also be sparse, to within mild changes in sparsity pattern and index due to the slight non-unity condition number. We therefore show in this section that the matrix of interest can be expected to be sparse, and subsequently introduce a drop tolerance technique in Sec. \ref{filter} to identify which nonzero entries to retain.

We note that $\Uc$ is comprised of eigenvectors of the 2D vector Laplacian, $\bC\bCt = \Uc \lambdatrue \Uc^T$, that mimics $\nabla^2 \velocity$. In Cartesian coordinates, $\nabla^2 \velocity$ reduces to the scalar Laplacian applied to each velocity component. Therefore, we can segregate $\Uc$ as $\Uc \equiv [\Ucx, \Ucy]^T$ where $\Ucx \in \RR^{\nqx \times \ns}$ and $\Ucy \in \RR^{\nqy \times \ns}$ are the eigenvectors of the scalar Laplacian acting on the $x$ and $y$ velocities, $u$ and $v$, respectively. On staggered grids, $u$ and $v$ have mixed boundary conditions on cell faces to enforce zero vorticity conditions on cell vertices. For instance, homogeneous Dirichlet boundary conditions in the $x$-direction and Neumann boundary conditions in the $y$-direction are imposed on the $u$-velocity and vice versa for $v$-velocity. Therefore, $\Uc$ contains a mixture of sines and cosines -- $\Ucx = \Sx \otimes \Cy$ and $\Ucy = \Cx \otimes \Sy$, where $\otimes$ denotes the Kronecker product, $\Sx$ and $\Sy$ are 1D discrete sine transforms (type-I) and $\Cx$ and $\Cy$ are 1D discrete cosine transforms (type-II, excluding the constant $[1,\ldots,1]^T$ vector that spans the null space of the Neumann operator). On substituting these decompositions we obtain
\begin{equation}
\Uc\Uc^T \equiv \begin{bmatrix}
\Sx\Sx^T \otimes \Cy\Cy^T & \Sx\Cx^T \otimes \Cy\Sy^T\\ 
\Cx\Sx^T \otimes \Sy\Cy^T & \Cx\Cx^T \otimes \Sy\Sy^T
\end{bmatrix}
\end{equation}
Here, the block diagonal entries are approximately identity because the sines and cosines are mutually orthogonal among themselves. Note that they are not \emph{exactly} identity because the discrete cosine vectors are truncated by one due to the exclusion of the constant null space vector. On the other hand, for the off-diagonal block terms, consider for instance, the continuous counterpart of the $(i,j+1)$ component of $\Sx\Cx^T$, 
\begin{equation}
(\Sx\Cx^T)_{(i,j+1)} = (\Sx^T\Cxiii)_{(i,j+1)} \xrightarrow[\text{analog}]{\text{continuous}} \int_0^L \sin\frac{i\pi x}{L} \cos\left(j+\frac{1}{2}\right)\frac{\pi x}{L} dx = \frac{L}{\pi(2i-2j-1)} + \frac{L}{\pi(2i+2j+1)}
\label{offdiag}
\end{equation}
\begin{figure}
\centering
\includegraphics[scale=1]{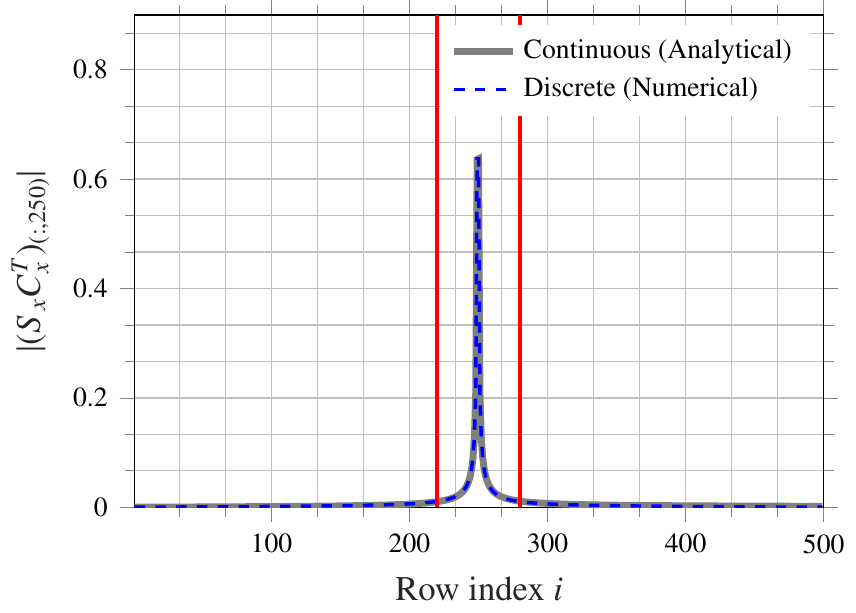}
\caption{Plot of the $250^{th}$ column of the discrete $|\Sx\Cx^T|$ obtained numerically via Fourier transforms and the continuous $|\Sx\Cx^T|$ obtained analytically from Eq. \eqref{offdiag}. The region between the red lines indicate the non-zero locations retained when applying a relative drop tolerance of $10^{-2}$ relative to $L/2$ for the continuous part and 1 for the discrete part.}
\label{sparsityfig}
\end{figure}
where $\Cxiii=\Cx^T$ is the discrete cosine transform of type-III \cite{strang1999discrete}. From Eq. \eqref{offdiag} it can be seen that any row or column of $|\Sx\Cx^T|$ is the discrete analog to a quantity that decays as $\frac{1}{|i-j|}$ with a peak when $i=j$. For reference, consider a problem with grid dimensions $500\times500$, for which we plot in Fig. \ref{sparsityfig} the $250^{th}$ column of the discrete $|\Sx\Cx^T|$ obtained numerically via Fourier transforms and the continuous $|\Sx\Cx^T|$ obtained analytically from Eq. \eqref{offdiag}. Note that for plotting the analytical part, the right most expression from Eq. \eqref{offdiag} is scaled (multiplied) by $1/(L/2)$, since $L/2$ is the value corresponding to the continuous counterpart of the diagonal of $\Sx\Sx^T$. The discrete $|(\Sx\Cx^T)_{(:,250)}|$ decays similarly to it's continuous counterpart as $\frac{1}{|i-j|}$.
The decay rate is increased as $\frac{1}{|i-j||k-l|}$ when we consider the entire off-diagonal block $\Sx\Cx^T \otimes \Cy\Sy^T$, where $k$ and $l$ indices correspond to the $(k+1,l)$ component of $\Cy\Sy^T$. Similar decaying trends can be derived for the remaining off-diagonal block $\Cx\Sx^T \otimes \Sy\Cy^T$. Under a drop tolerance filtering criteria where the matrix elements below a specified relative tolerance be dropped to zero, these off-diagonal blocks can be approximated sparsely. 

To indicate the impact of applying this drop-tolerance filtering procedure, a tolerance of $10^{-2}$ relative to $L/2$ will result in retaining approximately $60$ non-zeros per row for $\Sx\Cx^T$ irrespective of the size of the problem. For the illustration in Fig. \ref{sparsityfig}, the $\sim60$ non-zero locations retained for $|(\Sx\Cx^T)_{(:,250)}|$ are depicted by the region between the red lines. For a problem with grid dimensions $500 \times 500$ this leads to 8 times fewer kept entries per row than for the unfiltered case of 500 non-zeros. On accounting for the Kronecker products as well as the identity nature of the block diagonal entries, the fully filtered $\Uc\Uc^T$ will have 128 times fewer non-zeros compared to the unfiltered one. The savings, of course, will only increase with problem size---for example, rows or columns of $\Uc\Uc^T$ for a grid of dimensions $2500 \times 2500$ will have the same decay rate as the $500 \times 500$ case, and thus the same number of nonzero entries to be stored. 

We note that the above mentioned theoretical estimates of the sparsity pattern are based on two assumptions: (i) a constant diagonal matrix $\bIs + \alpha\lambdatrue$, and (ii) an equal segregation of $\Uc \equiv [\Ucx,\Ucy]^T$; \emph{i.e.}, for the $i^{th}$ column of $\Uc$, the associated discrete Fourier functions ${\Ucx}_{(i)}$ and ${\Ucy}_{(i)}$ are afforded equal weighting so that $||{\Ucx}_{(i)}||^2=||{\Ucy}_{(i)}||^2=0.5$. Regarding assumption (i), $\bIs + \alpha\lambdatrue$ is not a constant matrix but has a low condition number (as mentioned above), and therefore does not significantly alter the sparsity pattern of $\Uc\Uc^T$. Regarding assumption (ii), the non-equal weighting of the eigenvectors can be accounted for by incorporating diagonal matrices  ${D}_x$ and ${D}_y$ that unequally scale the different discrete Fourier functions: $\Uc = [\Ucx{D}_x,\Ucy{D}_y]^T$. This unequal weighting to the columns can be shown to only distribute the sparsity pattern across the diagonal and off-diagonal blocks, and not affect the overall number of non-zeros per row of $\Uc\Uc^T$. %Therefore, even in the real scenario where these assumptions are absent, $\bC(\bCt \bA \bC)^{-1}\bCt$ can be approximated sparsely and significant computational storage savings can be achieved.

Finally, returning to our original goal---we are interested in the overall sparsity of $\bB = \bEo\bC(\bCt \bA \bC)^{-1}\bCt\bEo^T$ instead of $\bC(\bCt \bA \bC)^{-1}\bCt$ alone. We note that $\bEo$ contains narrow delta functions (see Sec. \ref{interpolationEo}) that are discrete analogues to the Dirac delta function, and is therefore sparse with only a few nonzero entries off of each diagonal. Additionally, since $\bEo$ is a rectangular matrix, the overall size of $\bEo\bC(\bCt \bA \bC)^{-1}\bCt\bEo^T$ is further reduced, allowing for further efficiency gains in storing $\bB$.% which will have with approximately the same, if not reduced number of non-zeros per row as estimated.

%=======================================================================================
\subsubsection{Drop tolerance filtering technique}
\label{filter}

We now describe the drop tolerance filtering technique to construct $\bB$ sparsely. In this strategy, a drop tolerance parameter, $\droptol$, is used to filter out the elements of the matrix having relative magnitudes lower than the set tolerance. If we denote the sparsified version of $\bB$ as $\bB'$, then the filtering process is given as,
\begin{equation}
b'_{i,j} = \left\{\begin{matrix*}[l]
b_{ij} \quad \text{if} \ \ \ |b_{ij}|>\droptol \ |b_{ii}|\\ 
0 \quad \ \text{otherwise}
\end{matrix*}\right.
\label{filtereq}
\end{equation}
where $b_{ij}$ and $b'_{ij}$ are the $(i,j)^{th}$ element of $\bB$ and $\bB'$, respectively. Hereby, $\bB$ is replaced by the filtered matrix $\bB'$ in our proposed sub-domain based IB method. 

Since this filtering technique introduces additional approximations in the algorithm, the choice of $\droptol$ should be made judiciously. A large choice of $\droptol$ will proportionally filter out a large portion of $\bB$ and result in an unstable or inaccurate algorithm. On the other hand, a small $\droptol$ will yield only minimal storage gains. Through numerical testing, a drop tolerance of $\droptol=0.007$ is observed to strike the right balance between accuracy of the solutions and the storage requirements. This value of $\droptol$ is shown to be suitable for a variety of problems described in Sec. \ref{experiments}.

Finally, we emphasize that, in practice, we do not construct the full matrix $\bB$ before applying the filter. Instead, the columns of $\bB$ are constructed one at a time by successively computing the action of $\bB$ on a canonical unit vector as,
\begin{equation}
\left( \bEo\bC(\bCt \bA \bC)^{-1}\bCt\bEo^T \right) {e}_j = \bB_j
\label{action}
\end{equation}
where ${e}_j \in \RR^{2\nsd}$ is the $j^{th}$ canonical unit vector and $\bB_j$ is the $j^{th}$ column of $\bB$.
 The filter \eqref{filtereq} is then applied on $\bB_j$ before the next column, $\bB_{j+1}$, is evaluated. This construction process is conducive to scaling up for larger problem sizes.

%=======================================================================================

\subsection{Summary of the proposed sub-domain approach}

To summarize, the time-varying IB interpolation operator $\Enk{n+1}{k}$ defined on the moving Lagrangian body points is approximated via an interpolation of the time-independent IB interpolation operator $\Eo$ defined on a fixed set of Eulerian sub-domain points. This allows us to precompute $\B$ and circumvent the expensive $(\Ct \A \C)^{-1}$ solves traditionally required in Eq. \eqref{stress}. The full fractional step algorithm from Eq. \eqref{predictor}-\eqref{corrector} for our proposed sub-domain approach can be now written as,
\begin{equation}
\streamfnc^* = (\Ct \A \C)^{-1} \fluidrhs
\label{predictor2}
\end{equation}
\begin{equation}
\begin{split}
\left( \Pnk{n+1}{k} \B' \Ptnk{n+1}{k} + \frac{2\ds}{\dt} \Stnk{n+1}{k} \J^{-1} \Snk{n+1}{k} + \frac{2}{\dt} \hat{I}_{d}^T {\Jd}^{-1} \Sdnk{n+1}{k} \right) \stressnk{n+1}{k+1} = \\ \Pnk{n+1}{k}\Eo\C\streamfnc^* - \bcrhs{k} + \Stnk{n+1}{k} \left( \deflectionrhs{k} - \frac{2}{\dt}\J^{-1}\deflectiondotrhs{k} \right) +  \hat{I}_{d}^T \left( \defbodypointrhs{k} - \frac{2}{\dt}{\Jd}^{-1}\defbodypointdotrhs{k} \right)
\label{stress2}
\end{split}
\end{equation}
\begin{equation}
\deflectionincrease = \J^{-1} \left( \deflectiondotrhs{k} + \ds\Snk{n+1}{k} \stressnk{n+1}{k+1} \right)
\label{dtheta2}
\end{equation}
\begin{equation}
\defbodypointincrease = {\Jd}^{-1} \left( \defbodypointdotrhs{k} + \Sdnk{n+1}{k} \stressnk{n+1}{k+1} \right)
\label{dchi2}
\end{equation}
\begin{equation}
\streamfnc_{n+1} = \streamfnc^{*} - (\Ct\A\C)^{-1} \Ct\Eot\Ptnk{n+1}{k}\stressn{n+1}
\label{corrector2}
\end{equation}
Note that $\Enk{n+1}{k}$ in Eq. \eqref{predictor}--\eqref{corrector} is replaced by $\Pnk{n+1}{k}\Eo$ in Eq. \eqref{predictor2}--\eqref{corrector2} wherever applicable and the sparsified operator $\B'$ is used instead of $\B$ in Eq. \eqref{stress2}. %This system of equations now only requires two $(\Ct \A\C )^{-1}$ solves in equations \eqref{predictor2} and \eqref{corrector2} since $\B = \Eo \C (\Ct\A\C)^{-1}\Ct \Eot$ can be precomputed once and for all at the beginning of the first time step and stored sparsely. 

The entire sub-domain based IB method can be divided into offline and online stages. The offline stage is only performed once at the beginning of the simulation to compute $\B'$. In the online stage, the system of equations \eqref{predictor2}-\eqref{corrector2} are solved for the flow and structure variables and advanced in time. These stages are summarized in Algorithms \ref{algo1} and \ref{algo2}, respectively.

\begin{algorithm}
  \caption{Offline stage}\label{algo1}
  \begin{algorithmic}[1]
    \REQUIRE Problem setup and grid
    \ENSURE Precomputed and sparsified matrix $\bB'$
    \STATE Define a sub-domain according to the guidelines in Sec. \ref{choosesd1}
    \STATE Construct $\Eo$ using Eq. \eqref{ibsdinterp}
    \FOR{$j\leftarrow 1$ to $2\nsd$}
    \STATE Compute $j^{th}$ column ${\bB}_j$ from Eq. \eqref{action}
    \STATE Apply filtering: $\bB'_j \leftarrow filter(\bB_j)$ where $filter$ refers to the drop tolerance filtering technique in Eq. \eqref{filtereq}
    \ENDFOR
  \end{algorithmic}
\end{algorithm}
\begin{algorithm}
  \caption{Online stage}\label{algo2}
  \begin{algorithmic}[1]
    \REQUIRE Initial conditions $\streamfnc_{0}$, $\stressn{0}$ and $\bodypointn{0}$; precomputed matrix $\bB'$
    \ENSURE $\streamfnc_{n}$, $\stressn{n}$ and $\bodypointn{n}$ for $n=1,\ldots,t_{max}$
    \FOR{$n\leftarrow 0$ to $t_{max}$}
    \STATE Compute $\streamfnc^*$ from Eq. \eqref{predictor2}
    \STATE Initiate FSI iterations; $k\leftarrow 0$, $\stressnk{n+1}{0} = \stressn{n}$ and $\bodypointn{n+1}^{(0)} = \bodypointn{n}$
    \WHILE{$||\Delta \bodypointn{}||_{\infty}>\varepsilon$}
    \STATE Choose sub-domain points for interpolation based on Sec. \ref{choosesd} and construct interpolation matrix $\Pnk{n+1}{k}$ with weights from Eq. \eqref{weights}.
    \STATE Compute $\Pnk{n+1}{k} \B' \Ptnk{n+1}{k}$ sparsely and other structural operators 
    \STATE Solve Eq. \eqref{stress2} via GMRES for $\stressnk{n+1}{k+1}$
    \STATE Update position of the body $\bodypointn{n+1}^{(k+1)}$ via Eq. \eqref{dtheta2} and \eqref{dchi2}
    \STATE Advance FSI iterations $k\leftarrow k+1$%, $\stressnk{n+1}{k+2} = \stressnk{n+1}{k+1}$ and $\deflectionnk{n+1}{k+2} = \deflectionnk{n+1}{k+1}$
    \ENDWHILE
    \STATE Compute $\streamfnc_{n+1}$ from Eq. \eqref{corrector2}
    \ENDFOR
  \end{algorithmic}
\end{algorithm}

We note that the offline stage involves performing $(\Ct\A\C)^{-1}$ operations for every point in the sub-domain. Therefore, for a large and finely discretized sub-domain, precomputing $\B'$ can be an expensive process. However, we emphasize that it needs to be performed only once in the simulation. Furthermore, $\B'$ is independent of the instantaneous position of the bodies involved in the simulation. Therefore, $\B'$ constructed for a specific problem can be reused for several other problems provided that the following two conditions are met: (a) the spatial and temporal discretization sizes, Reynolds number and sub-domain remain unchanged and (b) all the bodies are guaranteed to stay within the sub-domain at all times. These conditions are conducive to parametric studies of flow problems, where only the body geometry or parameters such as mass ratio, stiffness \emph{etc.} are varied without modifying the underlying discretization or sub-domain. Therefore, such parametric studies, which are customary in the fluid dynamics community, can be efficiently performed using our proposed sub-domain based IB method.

%=======================================================================================

\section{Parallel implementation}
\label{parallelimplementation}

In this section, we describe the parallelization strategies implemented on the proposed sub-domain based IB approach to make it scalable across multiple CPUs.

%=======================================================================================

\subsection{Domain decomposition for fluid domain}

Domain decomposition is a technique used in parallel computing where the computational domain is partitioned among many processors and each processor solves a part of the same system of equations locally. During these local computations, any required information from the neighboring processors are communicated via a communication protocol. In this work we use the message passing interface (MPI) protocol. Domain partitioning is performed using the Portable, Extensible Toolkit for Scientific Computation (PETSc) \cite{petsc-user-ref} which is built using the MPI library. 

\begin{figure}
\centering
\includegraphics[scale=1]{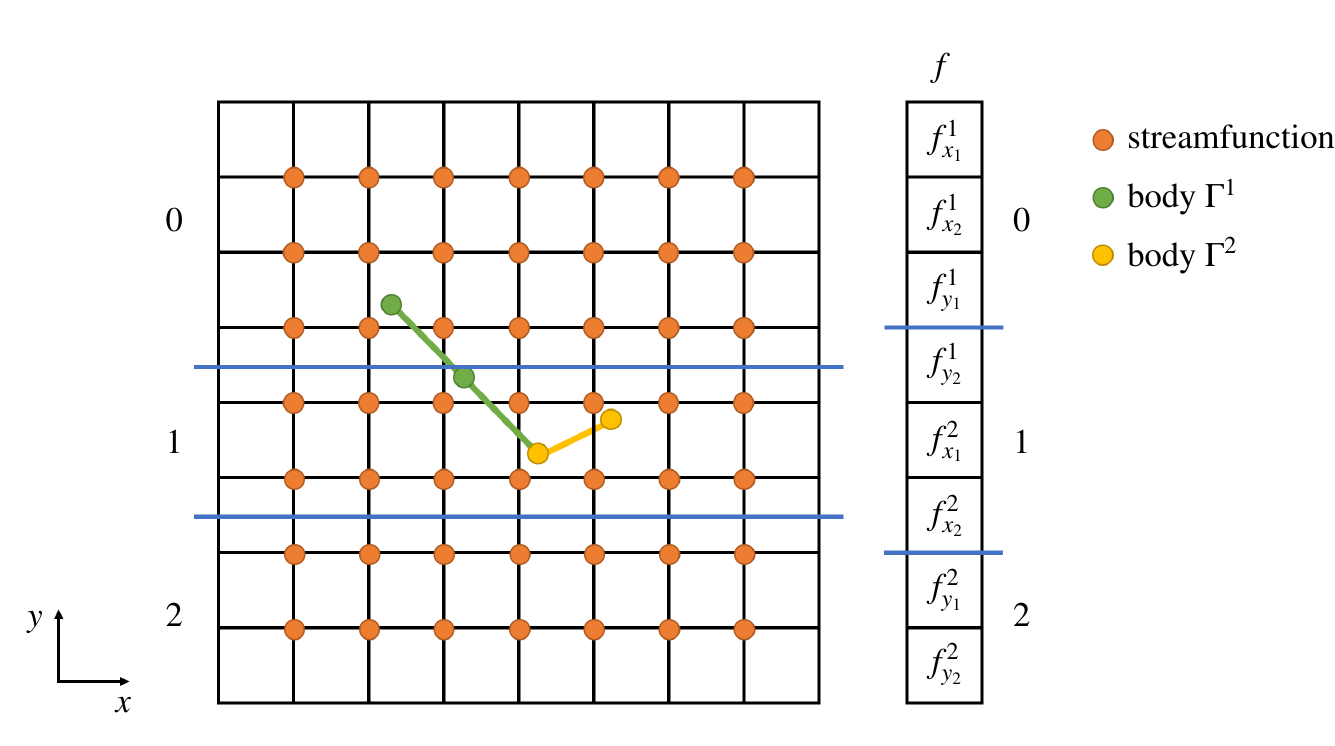}
\caption{Schematic of the domain decomposition of the fluid domain and partitioning of the surface stress vector among three processors labelled as 0, 1 and 2. The scripts in $\stressn{j_k}^i$ are, $i$: body number; $j$: $\{x,y\}$; $k$: body grid point. The blue lines denote the location of partitioning in the flow domain and surface stress vector.}
\label{parallel}
\end{figure}

The Poisson like operations involving $(\Ct \A \C)^{-1}$ are solved efficiently using fast sine transforms provided by the distributed-memory Fast Fourier Transform in the West (FFTW) MPI library \cite{FFTW05}. FFTW MPI requires that the domain be partitioned in only one dimension irrespective of a two or three dimensional flow domain. In Fortran, this partitioning is done along the last dimension of the domain; for instance, the $y$-direction for 2D problems and the $z$-direction for 3D. Fig. \ref{parallel} illustrates this domain partitioning procedure where the $y$-dimension is partitioned among three processors labelled as 0, 1 and 2. The blue lines in the flow domain denote the location of partitioning. Each processor handles the data computation involving the orange grid points in their respective domains. The inter-processor communication required while performing fast Fourier transforms is also managed by FFTW MPI.

As part of the domain partitioning technique, PETSc provides communication protocols conducive to the finite difference scheme used in our approach. Therefore, the inter-processor communications involved in operations such as $\C$ and $\Ct$ for computation at the grid points at the boundaries of the partitioned domains is efficiently handled by PETSc.

%=======================================================================================

\subsection{Partitioning of structure and flow sub-domain}

Eq. \eqref{stress2} is solved for the surface stress vector $\bstress \in \RR^{\nf}$ and the parallelization of Eq. \eqref{stress2} depends on the parallelization of $\bstress$. This vector consists of surface stresses in all coordinate directions for all the bodies involved in the simulation. 
In this work, we partition the entire surface stress vector $\bstress$ among a subset of available processors as equally as possible. We note that, since the number of degrees of freedom $\nf$ is very small compared to the flow grid points, over-partitioning $\bstress$ among a large number of processors can sometimes create a communication overhead which can result in negative scaling. Therefore, the choice of the number of subset processors is problem dependent. For all the problems considered in Sec. \ref{experiments}, we partition $\bstress$ among all the processors since we did not observe the aforementioned overhead. The partitioning procedure of the surface stress vector is also illustrated in Fig. \ref{parallel} where we consider a simple case of two bodies denoted by green and yellow points. We stack the surface stresses in the order of the number assigned to the body with stresses in $x$-direction stacked first followed by $y$-surface stress. This force vector is partitioned among three processors as denoted by the blue lines. Finally, Eq. \eqref{stress2} is solved in parallel using GMRES which is also provided by PETSc.

% for which there are several approaches in literature. In one of the approaches, the processor of the domain in which the Lagrangian body point exists, stores the $x$ and $y$ surface stresses corresponding to that body point. Since the body points can move from one domain to another in time, the processor assigned to that body point will also vary and therefore, a meticulous book-keeping is required.
%In a second class of approaches, a master-slave strategy is used where the entire Eq. \eqref{stress2} is solved in only one processor called as the `master', and the results of the computation are scattered to the other processors called as `slaves'. One of the disadvantages of these approaches are the lack of load-balancing of the number of degrees of freedom in the processors. 

The sub-domain and related operators are also partitioned similarly to the surface stress. For instance, the sub-domain IB interpolation operator $\bEo \in \RR^{2\nsd \times \nq}$ and the operator in Eq. \eqref{stress2} $\bB' \in \RR^{2\nsd \times 2\nsd}$ are partitioned equally along the first dimension \emph{i.e.} rows having a global dimension of $2\nsd$. The sparse interpolation operator $\bPnk{n+1}{k} \in \RR^{\nf \times 2\nsd}$ is also partitioned along the first dimension, but having a dimension $\nf$ and evaluated locally. %However, for an efficient computation of $\bPnk{n+1}{k} \bB' \bPtnk{n+1}{k}$, a redundant copy of the full operator $\Pnk{n+1}{k}$ is made available on every processor via inter-processor communication. This, however, does not create a communication overhead because the communication is overlapped with the computation of $\bEo$. Furthermore, since the total number of non-zeros in $\bPnk{n+1}{k}$ is $\nneighbors\times\nf=4\nf \ll \ns$ is small, it is not computationally prohibitive to store it redundantly on all processors even for very large problems.

%=======================================================================================

\subsection{Parallel interfacing between fluid and structure}

%The interfacing between fluid and structure is achieved via interpolation and regularization operators, $\Eo$ and $\Eot$, respectively.  
Although the above-mentioned flow domain and surface stress partitioning approaches ensure equal load-balancing across processors in their respective flow or structural domain, parallel interfacing between them is not trivial. For instance, consider the interpolation of velocity from the flow grid to the body points via $E \velocityq$, where $\velocityq \equiv \C \streamfnc$ is a generic velocity vector. Here, $\velocityq$ in the flow domain and $E$ of the body are partitioned via fundamentally different strategies. Therefore, to enable parallel interfacing, the velocity at flow points within the support of the delta function at the body point in consideration are ``scattered'' or communicated to the processor owning that body point. Once the scattering of the velocity data is performed, $E\velocityq$ can be trivially performed as a sparse matrix-vector multiplication. 

The exact same strategy is used for performing $\Eo \velocityq$ on the sub-domain in Eq. \eqref{stress2}. However, the size of $\Eo \velocityq$ is potentially much larger than $E \velocityq$, $\nsd \gg \nf$. Therefore, to improve the computational efficiency of performing $\Eo \velocityq$, it is evaluated at only those vector locations where the corresponding column of $\Pnk{n+1}{k}$ is non-zero since we eventually only need to evaluate the overall matrix-vector product $\Pnk{n+1}{k} \Eo \velocityq \ $.

%=======================================================================================

\section{Results}
\label{experiments}

In this section, we test the computational accuracy and efficiency of our proposed sub-domain based IB approach on several 2D FSI problems. Although our formulation in Sec. \ref{background} is developed for FSI problems involving arbitrary number of rigid, torsionally mounted and deformable bodies, for simplicity, the 2D problems considered in this section consist of a combination of rigid and torsional bodies.
The first problem consists of flapping of torsionally connected plates where we verify the accuracy of our sub-domain-based approach by comparing the results with those obtained by \citet{wang2015strongly} and using the true $\En{}, \Etn{}$ operators (Eq. \eqref{predictor}--\eqref{corrector}) in place of the sub-domain interpolation approximations (Eq. \eqref{predictor2}--\eqref{corrector2}). In the second problem, the use of a compact sub-domain is demonstrated on flow around a stationary airfoil with a passively deployable flap. 
The computational efficiency of our sub-domain approach is compared with that attained when using the true $\En{}, \Etn{}$ operators. These first two problems are constructed to highlight the accuracy and algorithmic efficiency of our proposed sub-domain-based interpolation approach. We then demonstrate the parallel scalability of our proposed method on a third problem consisting of 8 million grid points and increased complexity of a system of three airfoils in tandem each equipped with three passively deployable flaps. 

A multi-domain approach for far-field Dirichlet boundary conditions of zero vorticity is incorporated for solving the flow equations where a hierarchy of grids of increasing coarseness stretching to the far field is employed (see reference \cite{colonius2008fast} for details). Following \citet{goza2017strongly}, the immersed boundary spacing is set to be twice as that of the flow grid spacing of the finest grid. A convergence criteria of $\| \deflectionincrease \|_{\infty} \le 10^{-7}$ is used when iterating between Eq. \eqref{stress2} and \eqref{dtheta2}. The relative error used in various grid convergence and comparison studies in this section is defined as,
\begin{equation}
\text{Error} (\%) = \frac{||\eta-\eta_{ref}||_2}{||\eta_{ref}||_2} \times 100
\label{errortheta}
\end{equation}
where $\eta$ is the quantity of interest compared against a reference $\eta_{ref}$.

%=======================================================================================

\subsection{Flapping of torsionally connected plates}
\label{eldg}

\subsubsection{Problem description}

\begin{figure}
\centering
\includegraphics[scale=1]{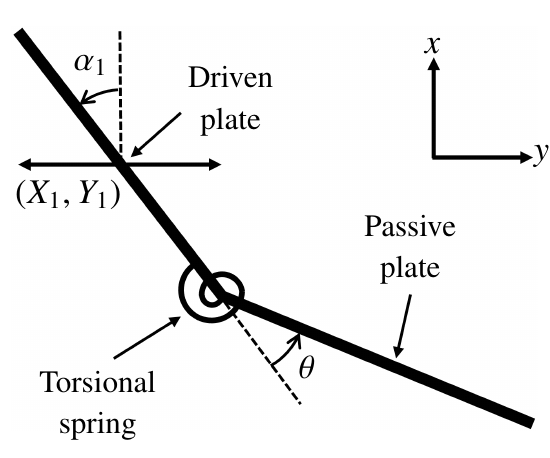}
\caption{Schematic of the flapping of two plates connected by a torsional spring.}
\label{schematiceldg}
\end{figure}

This problem involves flapping of a 2D wing modeled in \citet{wang2015strongly}. In reference \cite{wang2015strongly}, the wing was modeled as two ellipses of chord length $c$ having aspect ratios of 5:1, connected via a torsional spring. However, for simplicity, we model the ellipses as flat plates due to the high aspect ratio of the ellipses. A schematic of this problem is shown in Fig \ref{schematiceldg}: a `driven' plate oscillates according to prescribed kinematics, and a second plate that is hinged at one end of the driven plate undergoes dynamics determined by the balance of aerodynamic and structural (stiffness and inertial) forces. The dimensional equation of motion for the hinge deflection angle $\deflection$ between the plates is given by,
\begin{equation}
I \ddot{\deflection} + C \dot{\deflection} + K \deflection = M_f -[m L_2 \cos(\alpha_1 + \deflection)]\ddot{X}_1 - [\inertiadim + m(L_2^2 + L_1L_2\cos\deflection)]\ddot{\alpha}_1 - mL_1L_2\sin\deflection\dot{\alpha}_1^2
\label{newtonseldg}
\end{equation}
where $m$ is the mass of the passive plate, $M_f$ is the dimensional moment due to the aerodynamic body forces analogous to the integral term in Eq. \ref{newtons} and $L_1=L_2=0.55c$ are the distances from the center of gravity of the respective plates to the hinge. Although the bodies are separated by a gap of width $0.1c$ in \cite{wang2015strongly}, we neglect the gap and extend both the plates up to the hinge since the gap has negligible effect on the aerodynamics \cite{toomey2009numerical}. The length of the plates is therefore set to $1.05$c.
The stiffness and damping coefficient of the spring are $K/(\densityf f^2 c^4) = 456$ and $C/(\densityf f c^4) = 3.95$, respectively, where $f$ is the frequency of oscillations of the driven plate. The moment of inertia of the plate is $I/(\densityf c^4) = 0.2886$ which is equivalent to a density ratio of $\rho_s/\densityf=5$ for the ellipse in \cite{wang2015strongly}. The multi-domain approach for far-field boundary conditions uses 5 grids of increasing coarseness where the finest and coarsest grid levels are $[-3.15,3.15]c \times [-4.65, 1.65]c$ and $[-50.4,50.4]c \times [-51.9, 48.9]c$, respectively. 

%-------------------------------------
%-------------------------------------

The kinematics prescribed on the driven plate are same as that were used in \citet{wang2015strongly}, given by
\begin{equation}
X_1(t) = \frac{A_0}{2}\frac{G_t(ft)}{\max G_t} C(ft)
\end{equation}
\begin{equation}
Y_1(t) = 0
\end{equation}
\begin{equation}
\alpha_1(t) = -\beta\frac{G_r(ft)}{\max G_r}
\end{equation}
where the translational and rotational shape functions, $G_t(t)$ and $G_r(t)$, respectively are given by,
\begin{equation}
G_t(t) = \int_t \tanh[\sigma_t \cos(2\pi t')]dt'
\end{equation}
\begin{equation}
G_r(t) = \tanh[\sigma_r \cos(2\pi t)]
\end{equation}
The initial impulsive velocity is avoided by using a start-up conditioner given by,
\begin{equation}
C(t) = \frac{\tanh(8t-2)+\tanh 2}{1+\tanh 2}
\end{equation}
Based on these kinematic parameters, the rotational Reynolds number is defined as,
\begin{equation}
Re_r = \frac{2\pi \beta \sigma_r}{\tanh \sigma_r} \frac{f c^2}{\nu}
\end{equation}
We consider two test cases corresponding to the kinematic parameters provided in Table \ref{kineparams}. See reference \cite{toomey2008numerical} for a detailed study of these parameters on the physics and aerodynamics of flapping. 

To set the boundaries of the rectangular sub-domain for our proposed approach, firstly we determine the maximum limits of the body displacements. The maximum $y$-limits of the body displacements are $[-1.6,0.5]c$ which may occur when  $\alpha_1=0$ and $\deflection=0$. For the $x$-limits, although the maximum body displacements are $[-1.99,1.99]c$ based on $\max(\alpha_1) = \beta = \pi/4$ and $\max(X_1) = A_0/2c = 0.7$, these maximum conditions never occur simultaneously since they are separated by a $\pi/2$ phase difference. Based on these conditions, the sub-domain is set to $[-1.89,1.89]c \times [-1.66, 0.65]c$ which is a conservative estimate of the maximum limits of the body displacements.

%-------------------------------------
\begin{table}
\centering
\begin{tabular}{|c|c|c|c|c|c|}
\hline
Case No. & $A_0/c$ & $\beta$ & $\sigma_t$ & $\sigma_r$ & $Re_r$ \\
\hline
1 & 1.4 & $\pi/4$ & 3.770 & 3.770 & 100 \\
\hline
2 & 1.4 & $\pi/4$ & 0.628 & 0.628 & 100 \\
\hline
\end{tabular}
\caption{Kinematic parameters for the flow problem of flapping of torsionally connected plates.}
\label{kineparams}
\end{table}
%-------------------------------------

%=======================================================================================

\subsubsection{Implementation}

%---------------------------------------
\begin{table}
\centering
\begin{tabular}{|l|c|c|c|}
\hline
$\dx/c$ & $\dt/\tau_r$ & Discrepancy in $\deflection$ & Discrepancy in $c_l$ \\
\hline
0.00525 & 0.00244 & & \\
 \hline
 0.0105 & 0.00489 & 0.56\% & 2.10\% \\
 \hline
 0.021 & 0.00978 & 1.82\% & 5.82\% \\
 \hline
 0.042 & 0.0196 & 8.26\% & 19.83\% \\
 \hline
\end{tabular}
\caption{Parameters for grid convergence study and corresponding discrepancies in $\deflection$ and $c_l$ reported with respect to the finest case of $\dx/c=0.00525$ for the problem of flapping of torsionally connected plates.}
\label{eldggc}
\end{table}
%----------------------------------------------
\begin{figure}
\centering
\begin{subfigure}[t]{0.48\textwidth}
\centering
\includegraphics[scale=1]{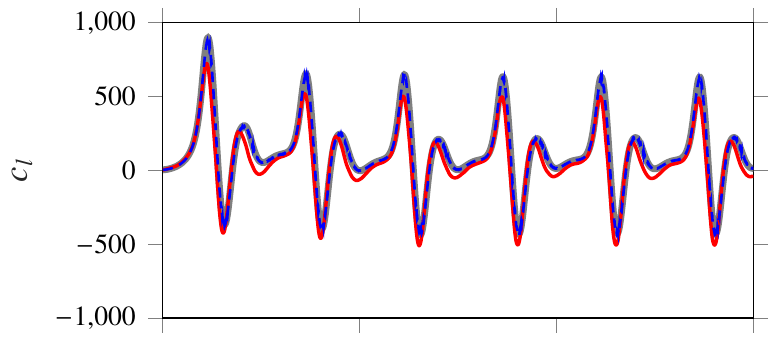}
\put(-140,115){\fbox{%
\color{gray}\hdashrule[0.5ex][x]{0.1\textwidth}{2pt}{} \color{black} \footnotesize True $\En{}, \Etn{}$ 
\quad \quad
\color{red}\hdashrule[0.5ex][x]{0.1\textwidth}{2pt}{} \color{black} \footnotesize \citet{wang2015strongly}
\quad \quad
\color{blue}\hdashrule[0.5ex][x]{0.12\textwidth}{1.5pt}{3pt} \color{black} \footnotesize Sub-domain approximation
}}
\caption{Case 1: $c_l$}
\end{subfigure}
\begin{subfigure}[t]{0.48\textwidth}
\centering
\includegraphics[scale=1]{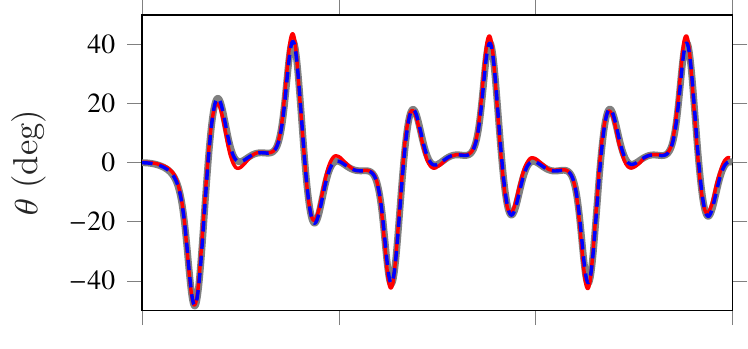}
\caption{Case 1: $\deflection$}
\end{subfigure}
\centering
\begin{subfigure}[t]{0.48\textwidth}
\centering
\includegraphics[scale=1]{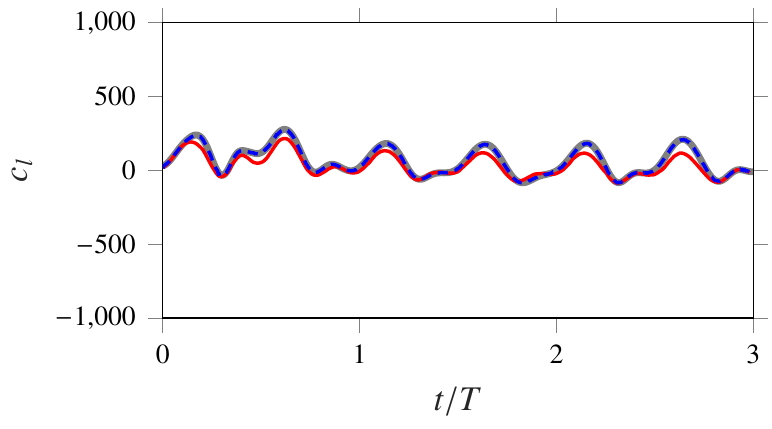}
\caption{Case 2: $c_l$}
\end{subfigure}
\begin{subfigure}[t]{0.48\textwidth}
\centering
\includegraphics[scale=1]{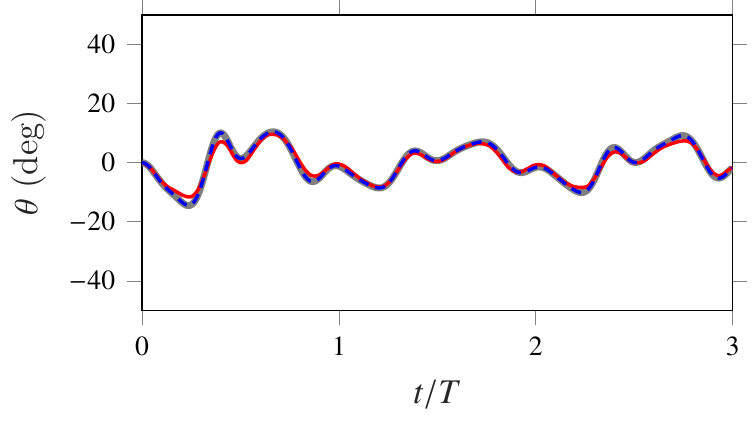}
\caption{Case 2: $\deflection$}
\end{subfigure}
\caption{Plots of lift coefficient, $c_l$ and deflection angle, $\deflection$ for the two cases obtained by Goza \emph{et al} \cite{goza2017strongly}, Wang \emph{et al} \cite{wang2015strongly} and our present approach for the problem of flapping of torsionally connected plate.}
\label{eldgplot}
\end{figure}
%----------------------------------------------
%----------------------------------------------
\begin{figure}
\centering
\begin{subfigure}[t]{0.32\textwidth}
\centering
\includegraphics[scale=1]{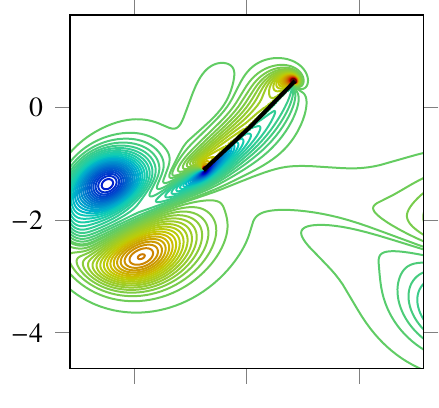}
\caption{$t/T=1.14$}
\end{subfigure}
\begin{subfigure}[t]{0.32\textwidth}
\centering
\includegraphics[scale=1]{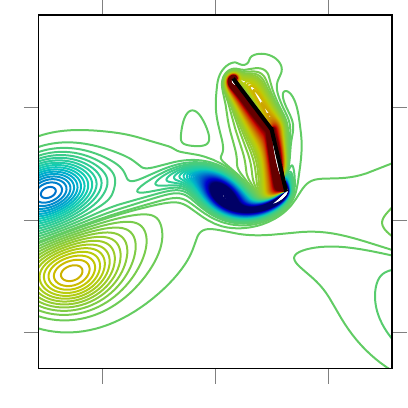}
\caption{$t/T=1.30$}
\end{subfigure}
\begin{subfigure}[t]{0.32\textwidth}
\centering
\includegraphics[scale=1]{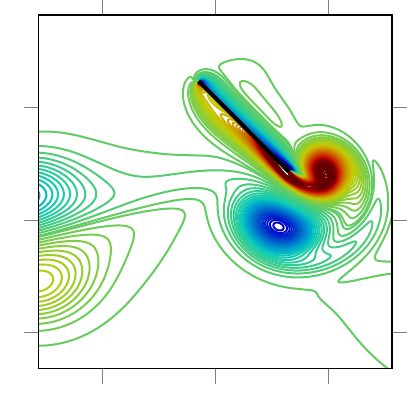}
\caption{$t/T=1.47$}
\end{subfigure}
\centering
\begin{subfigure}[t]{0.32\textwidth}
\centering
\includegraphics[scale=1]{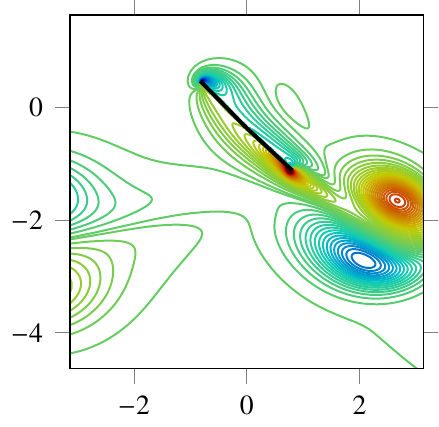}
\caption{$t/T=1.63$}
\end{subfigure}
\begin{subfigure}[t]{0.32\textwidth}
\centering
\includegraphics[scale=1]{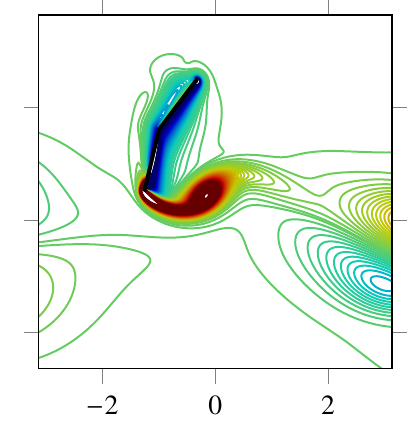}
\caption{$t/T=1.80$}
\end{subfigure}
\begin{subfigure}[t]{0.32\textwidth}
\centering
\includegraphics[scale=1]{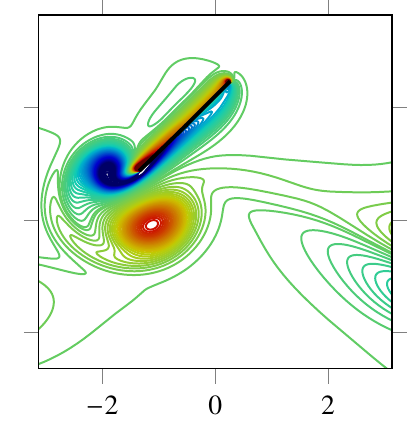}
\caption{$t/T=1.96$}
\end{subfigure}
\caption{Contour plots of vorticity at different time instants for case 1 of the problem of flapping of torsionally connected plates.}
\label{eldgsnapshots}
\end{figure}
%----------------------------------------------

Firstly, a grid convergence study on the first test case is performed by varying the spatial and temporal discretizations of the finest domain, $\dx/c$ and $\dt/\tau_r$, respectively, as shown in Table \ref{eldggc}, where $\tau_r = (2\pi \beta \sigma_r f/\tanh \sigma_r)^{-1}$ is the characteristic rotation time. The discrepancy in the deflection angle, $\deflection(t)$, and lift coefficient, $c_l(t)=2F_y(t)/\densityf^3c^3$, in $0<t/T<3$ computed using Eq. \eqref{errortheta} are used for determining convergence, where $T=f^{-1}$ is the time period and $F_y$ is the total force on both plates in the $y$-direction. In this grid convergence study, the finest grid with $\dx/c=0.00525$ is set to be the reference case against which the changes in deflection angle and lift are evaluated. Since the grid with $\dx/c=0.0105$ is converged to within 1\% of the finest grid for $\deflection$ as shown in Table \ref{eldggc}, $\dx/c=0.0105$ and $\dt/\tau_r=0.00489$ are used for presenting the results. Next, the order of spatial convergence $p$ is determined via Richardson extrapolation as,
\begin{equation}
p = \log\left( \frac{|\eta_{r^2\dx}-\eta_{r\dx}|}{|\eta_{r\dx}-\eta_{\dx}|} \right) /\log(r)
\end{equation}
where $\eta$ is a flow metric evaluated for successively refined grids with a constant refinement ratio of $r$ and subscript denotes the relative grid under consideration. In this problem, we set $\eta \equiv \deflection(t)$ and $r=2$. By using the first three grids in Table \ref{eldggc} and averaging $p$ in $0<t/T<3$, we get the spatial order of accuracy to be $p=1.34$. This is in agreement with the order of accuracy of most IB methods of between first and second order \cite{taira2007immersed}.

Next, we probe the accuracy of our proposed sub-domain approach by comparing the lift coefficient and deflection angle in Fig. \ref{eldgplot}, for the two test cases listed in Table \ref{kineparams}, to those obtained by \citet{wang2015strongly} and when using the true $\En{}, \Etn{}$ in place of the sub-domain interpolation approximations. The temporal variation of the deflection angle agrees well across all three cases, though the two approaches considered here have slight differences from the results of \citet{wang2015strongly} because we model the ellipses as flat plates. For completeness, we illustrate the passive flapping of the second plate and the resulting lingering vortices via vorticity snapshots at different time instants in Fig. \ref{eldgsnapshots}.

We provide the relative errors in the lift and deflection angle between our sub-domain interpolation approach and the use of the true $\En{}$, $\Etn{}$ operators in Table \ref{eldgerror}. Relative errors of less than 1\% and 2\% for the lift and deflection angle, respectively, are obtained, which are within the tolerance to which our results are converged; c.f., Table \ref{eldggc}.  For all the cases considered, a maximum of three FSI iterations were required per time step. The computational efficiency of our approach is also demonstrated in Table \ref{eldgerror} via the significant speed-up obtained by our sub-domain approach compared with use of the true $\En{}, \Etn{}$. Here, speed-up is defined as the ratio of mean wall-times incurred per time step in $0<t/T<1$ when the simulations are performed on a single core. The speed-up of an order of magnitude is due to the elimination of the bottleneck described in Sec. \ref{bottleneck}.

%----------------------------------------------
\begin{table}
\centering
\begin{tabular}{|c|c|c|c|}
\hline
Case & Error in $\deflection(t)$ & Error in $c_l(t)$ & Speed-up \\
 \hline
1 & 0.42\% & 1.65\% & 12.55 \\
\hline
2 & 0.49\% & 1.91\% & 10.17 \\
\hline
\end{tabular}
\caption{Demonstration of computational accuracy via relative errors in $\deflection$ and $c_l$, and speed-ups in using the proposed sub-domain approach compared to the true $\En{}, \Etn{}$ operators for the problem of flapping of torsionally connected plates.}
\label{eldgerror}
\end{table}

%=======================================================================================

\subsection{Passively deployed flap on an airfoil}
\label{airfoilproblem}

\subsubsection{Problem description}

\begin{figure}
\centering
\includegraphics[scale=1]{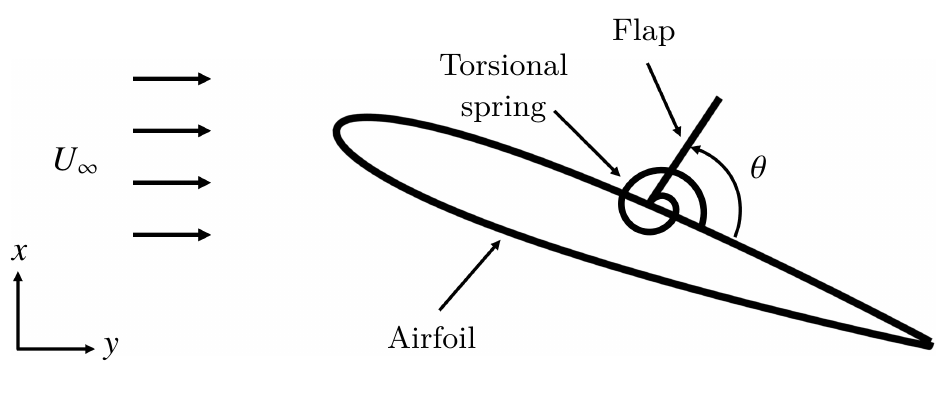}
\caption{Schematic of the system of passively deployable flap on an airfoil}
\label{schematicairfoil}
\end{figure}

This problem consists of a stationary NACA0012 airfoil of chord length $c$ at an angle of attack of $20^\circ$ in a flow with freestream velocity $\velocityscale$. The Reynolds number based on the chord length is set to 1000. A flap of length $0.2c$ is hinged on the upper surface of the airfoil at a distance of $0.5c$ from the leading edge via a torsional spring, as shown in Fig. \ref{schematicairfoil}. We fix the non-dimensional moment of inertia and damping coefficient to $\inertia= \inertiadim/\densityf c^4 =0.001$ and $\damper=\damperdim/\densityf\velocityscale c^3=0$, respectively and consider three test cases of widely varying stiffness, $\stiffness=\stiffnessdim/\densityf \velocityscale^2 c^2=\{0, 0.001, 0.1\}$. 
Initially, the flap is rested at an angle of $5^\circ$ from the airfoil surface, which is taken as the undeformed (zero stress) deflection angle. As the vortex shedding process occurs, the flap passively deploys and interacts with the flow, providing significant lift improvements compared to the flap-less case \cite{rosti2018passive,duan2018design}. For the multi-domain approach for far-field boundary conditions,  five grids of increasing coarseness are used where the finest and coarsest grid levels are $[-0.5,2.5]c \times [-1.5, 1.5]c$ and $[-23,25]c \times [-24, 24]c$, respectively. 

We chose the airfoil-flap problem to demonstrate the use of a compact sub-domain to reduce the storage requirements of the precomputed matrix $\bB'$. Since the airfoil is stationary and only the flap undergoes large displacements, we construct a small rectangular sub-domain that bounds only the physical limits of flap displacements. Accordingly, the rectangular sub-domain is set to $[0.23,0.7]c \times [-0.24, 0.1]c$. Now, to account for the stationary airfoil, the exact airfoil body points are appended into the set of sub-domain points. These exact body points also allow us to use the exact IB interpolation operator $\En{}$ for the airfoil by setting the interpolation weight to one in $\Pn{}$ corresponding to the airfoil points. In problems such as these where physical knowledge of the problem is available that yield a compact sub-domain, significant savings in storing the precomputed matrix $\bB'$ can be achieved. Finally, we emphasize that, since the underlying discretization sizes, sub-domain and $\reynolds$ are fixed, the precomputed matrix is only computed once for all the parametric variations considered within this test problem.

%=======================================================================================

\subsubsection{Implementation}
\label{airfoilimplementation}

%-------------------------------
\begin{table}
\centering
\begin{tabular}{|l|c|c|c|}
\hline
$\dx/c$ & $\dt/(c/\velocityscale)$ & Mean deflection $\bar{\deflection}$ & Discrepancy in $\deflection$ (\%) \\
\hline
0.0025 & 0.0003125 & $79.18^\circ$ & \\
 \hline
0.003 & 0.000375 &  $79.60^\circ$ & 0.53 \\
 \hline
0.00349 & 0.0004375 & $79.96^\circ$ & 0.99 \\
 \hline
0.00395 & 0.0004935 & $77.93^\circ$ & 1.58  \\
 \hline
0.00455 & 0.000568 & $76.92^\circ$ & 2.85  \\
 \hline 
\end{tabular}
\caption{Grid convergence test cases and corresponding errors in $\deflection$ reported with respect to the finest case of $\dx/c=0.0025$ for the problem of passively deployable flap on an airfoil.}
\label{airfoilgc}
\end{table}
%------------------------------------------------------
\begin{figure}
\centering
\begin{subfigure}[t]{0.46\textwidth}
\centering
\includegraphics[scale=1]{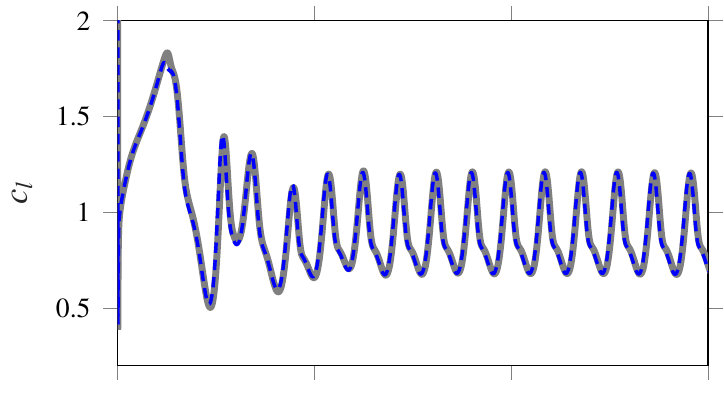}
\put(-50,132){\fbox{%
\color{gray}\hdashrule[0.5ex][x]{0.1\textwidth}{2pt}{} \color{black} \footnotesize True $\En{}$, $\Etn{}$ 
\quad \quad
\color{blue}\hdashrule[0.5ex][x]{0.12\textwidth}{1.5pt}{3pt} \color{black} \footnotesize Present
}}
\caption{Case 1: $c_l$}
\end{subfigure}
\begin{subfigure}[t]{0.46\textwidth}
\centering
\includegraphics[scale=1]{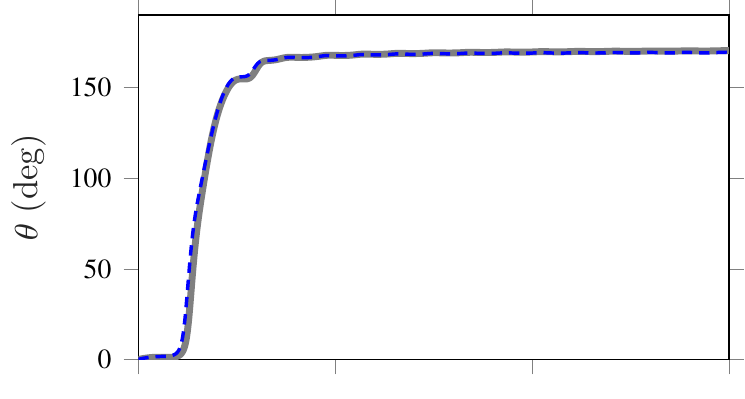}
\caption{Case 1: $\deflection$}
\end{subfigure}
\centering
\begin{subfigure}[t]{0.46\textwidth}
\centering
\includegraphics[scale=1]{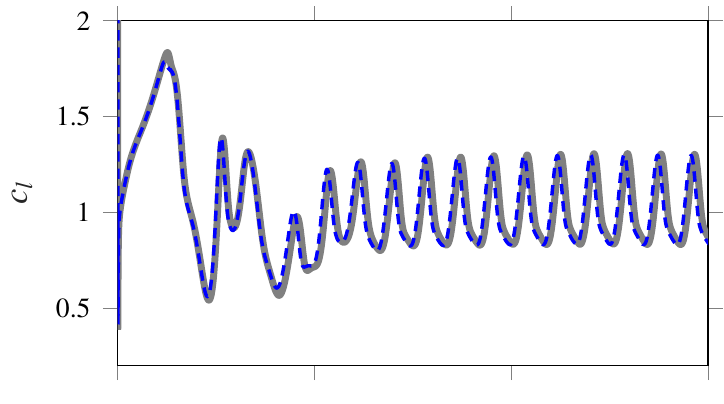}
\caption{Case 2: $c_l$}
\end{subfigure}
\begin{subfigure}[t]{0.46\textwidth}
\centering
\includegraphics[scale=1]{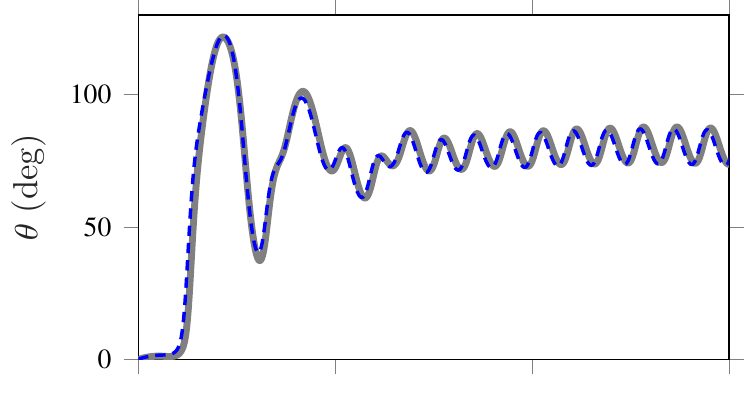}
\caption{Case 2: $\deflection$}
\end{subfigure}
\centering
\begin{subfigure}[t]{0.46\textwidth}
\centering
\includegraphics[scale=1]{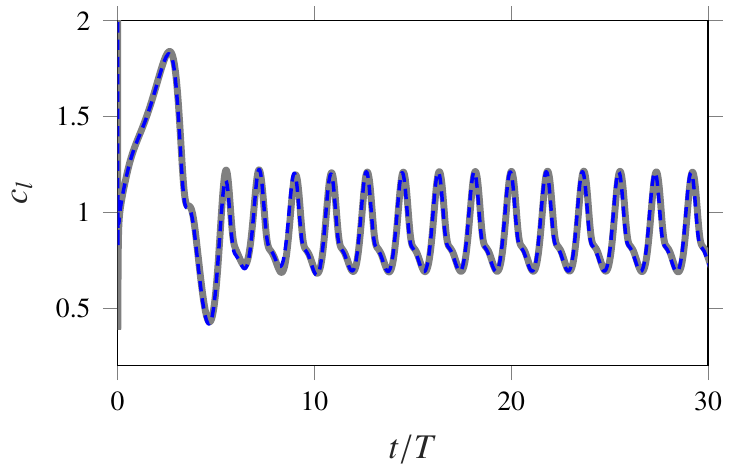}
\caption{Case 3: $c_l$}
\end{subfigure}
\begin{subfigure}[t]{0.46\textwidth}
\centering
\includegraphics[scale=1]{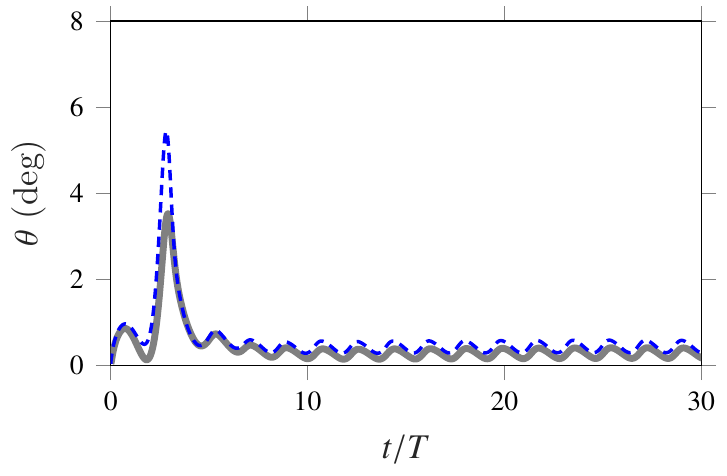}
\caption{Case 3: $\deflection$}
\end{subfigure}
\caption{Plots of lift coefficient, $c_l$ and deflection angle, $\deflection$ for the three cases of stiffness, $\stiffness=\{0,0.001,0.1\}$, obtained by Goza \emph{et al} \cite{goza2017strongly} and our present approach for the airfoil-flap system.}
\label{airfoilplots}
\end{figure}
%-------------------------------

Firstly, a grid convergence study on the test case of $\stiffness=0.001$ is performed by varying the spatial and temporal discretizations of the finest domain as shown in Table \ref{airfoilgc}. The mean deflection angle $\bar{\deflection}$ in the limit cycle oscillation regime ($t/(c/\velocityscale)>20$) is used to determine grid convergence. In this grid convergence study, the finest  grid with $\dx/c=0.0025$ is set to be the reference case against which the relative changes of mean deflection angle are computed. Since the grid with $\dx/c=0.00349$ is converged to within 1\% of the finest grid, $\dx/c=0.00349$ and $\dt/(c/\velocityscale)=0.0004375$ are used for presenting the results.

%-----------------------------------

Next, we determine the accuracy of our proposed sub-domain approach by comparing the lift coefficient and deflection angle in Fig. \ref{airfoilplots}, for the various cases of stiffness, $\stiffness=\{0, 0.001, 0.1\}$, to those obtained when using the true $\En{}$, $\Etn{}$ operators in place of the sub-domain interpolation approximations. Here, the lift coefficient is defined as $c_l=2F_y/\densityf\velocityscale^2c$  where $F_y$ is the total force on the airfoil and flap system in the $y$-direction. It can be seen that the transient dynamics of the flap deploying into the flow (implied from the large initial deflection angles) and subsequent limit cycle oscillations produced from our sub-domain approach agree well with those obtained by using true $\En{}$, $\Etn{}$ for all the cases. The plots of deflection angle also demonstrate the stability of our approach in the presence of large deflections for very low stiffness of $\stiffness=0$ and $\stiffness=0.001$. For all the cases considered, a maximum of only two FSI iterations were required per time step. The relative errors in the mean lift coefficient and deflection angle between our approach and the use of true $\En{}$, $\Etn{}$ are also provided in Table \ref{airfoilerror}. For all the cases, relative errors of less than 1\% for both the lift and deflection angle are attained. Note that, for the case of $\stiffness=0.1$, we have reported the absolute error in the mean deflection angle instead of the relative error because the flap oscillates very close to the airfoil with a mean deflection angle of $0.44^\circ$ and $0.29^\circ$ obtained from our sub-domain approach and by using true $\En{}$, $\Etn{}$, respectively. This results in a misleadingly high relative error of $52.79\%$ with respect to such a small mean deflection angle while noting that the relative error in $c_l$ is still below 1\%.

The computational efficiency of our approach is demonstrated in Table \ref{airfoilerror} by reporting the speed-up attained by our proposed approach compared to when the true $\En{}, \Etn{}$ operators are utilized. Here, the speed-up is defined as the ratio of mean wall-times on a single core incurred per time step  over the first 1000 time steps ($t/(c/\velocityscale)<0.4375$). Our proposed sub-domain approach is approximately four times more efficient than when using the true $\En{}, \Etn{}$ operators for this airfoil-flap problem.
%----------------------------------------
\begin{table}
\centering
\begin{tabular}{|c|c|c|c|c|}
\hline
$\stiffness$ & Error in $\bar{\deflection}$ & Error in $\bar{c}_l$ & Speed-up & $\Delta \bar{c}_l$ \\
 \hline
0 & 0.47\% & 0.41\% & 4.22 & -0.25\% \\
\hline
0.001 & 0.83\% & 0.30\% & 3.87 & 14.55\% \\
\hline
0.1 & $0.15^\circ$ & 0.48\% & 4.07 & 1.15\% \\
\hline
\end{tabular}
\caption{Results from the airfoil-flap problem: columns 2-3: demonstration of computational accuracy via relative errors in $\deflection$ and $c_l$, and efficiency via speed-ups with respect to the true $\En{}, \Etn{}$ operators; column 4: change in mean lift compared with a baseline case involving no flaps. The last column is not a measure of computational accuracy, but a demonstration of the potential aerodynamic benefits associated with torsionally-hinged flaps.}
\label{airfoilerror}
\end{table}
%---------------------------------------
%------------------------------------------------------
\begin{figure}
\centering
\begin{subfigure}[t]{0.46\textwidth}
\centering
\includegraphics[scale=1]{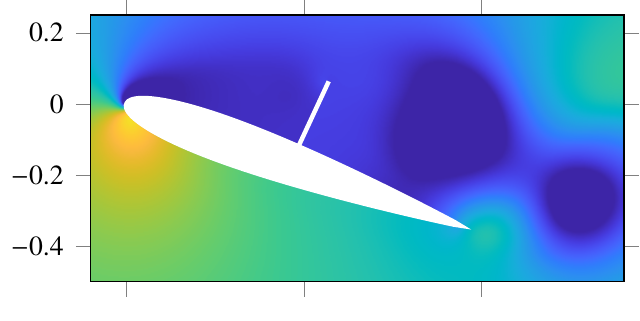}
\caption{$t/T=0$}
\end{subfigure}
\begin{subfigure}[t]{0.46\textwidth}
\centering
\includegraphics[scale=1]{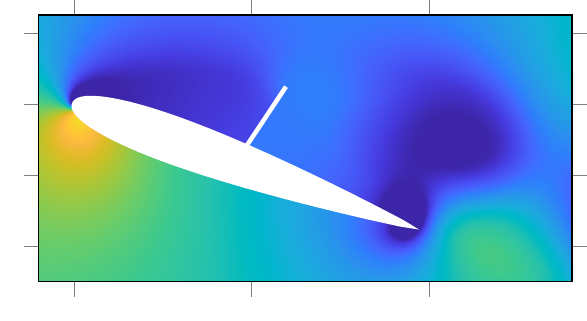}
\caption{$t/T=0.25$}
\end{subfigure}
\centering
\begin{subfigure}[t]{0.46\textwidth}
\centering
\includegraphics[scale=1]{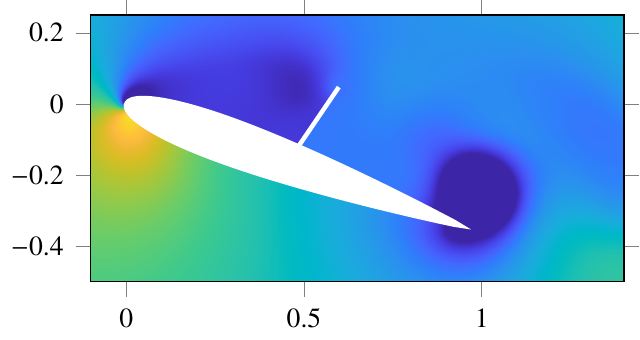}
\caption{$t/T=0.5$}
\end{subfigure}
\begin{subfigure}[t]{0.46\textwidth}
\centering
\includegraphics[scale=1]{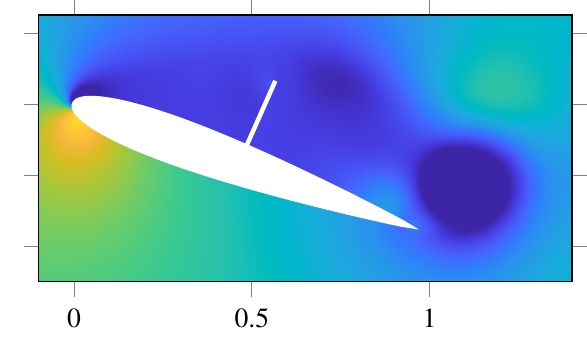}
\caption{$t/T=0.75$}
\end{subfigure}
\caption{Contour plots of pressure at four time instants in one time period $T$. Blue and yellow color denotes regions of low and high pressure, respectively.}
\label{airfoilsnapshots}
\end{figure}
%------------------------------------------------------

Finally, to indicate the potential engineering utility of these deployable flaps in improving aerodynamic performance, we show in Table \ref{airfoilerror} the relative change in the lift coefficient, $\Delta\bar{c}_l$, for the airfoil-flap system  compared with the flap-less case of only the airfoil at the same angle of attack and $\reynolds$. The case with $\stiffness=0.001$ provides significant lift benefits of around 15\%. To understand the physical mechanisms that enable this lift improvement, four snapshots of the pressure field over one period of the limit cycle oscillation regime ($t/(c/\velocityscale)>20$) are plotted in Fig. \ref{airfoilsnapshots}. We can clearly observe a low pressure region denoted by blue color just upstream of the flap in all the contours. This low pressure zone is formed due to the trapping of a portion of the leading edge vortex by the flap. This low pressure region therefore augments the lift of the airfoil-flap system compared to the case without the flap. Similar physical mechanisms that augment lift have been found for statically deployed flaps \cite{meyer2007separation}, but to our knowledge this mechanism has not been observed for the case of dynamic flaps mounted via torsional springs. The lift variations for $\stiffness=0$ and $\stiffness=0.1$ are not significant since they either excessively or barely deploy the flap, respectively, such that the trapping of the vortex is not realized. 

%=======================================================================================

\subsection{Airfoils with passively deployed flaps in tandem}
\label{tandemproblem}
\subsubsection{Problem description}

In this section, we demonstrate the parallel scalability of our proposed approach on a relatively large problem consisting of 8 million flow grid points.
This problem involves a similar airfoil-flap system as described in the previous problem in Sec. \ref{airfoilproblem}, but with three stationary NACA0012 airfoils in tandem, each equipped with three torsionally hinged flaps. Such a tandem-airfoil-flap system is found to reduce the total drag coefficient compared to the tandem-airfoil system without any flaps (see the next Sec. \ref{tandemimplementation} for details). 

The airfoils are separated by a distance of $1.18c$ between the consecutive leading edges where $c$ denotes the chord length of the airfoils. The flaps are located at a distance of $0.25c$, $0.5c$ and $0.75c$ from the leading edge of their respective airfoils. The angle of attack of all the airfoils is $20^\circ$ and Reynolds number of the flow based on $c$ is set to 1000. The parameters for all the springs and flaps are $\stiffness=0.001$, $\damper=0$ and $\inertia=0.001$. Initially, all the flaps are rested at an angle of $5^\circ$ from their respective airfoil tangential surface. As the vortex shedding process occurs, all flaps are allowed to passively respond to the aerodynamic forces. 

The multi-domain approach for the far-field boundary conditions employs five grids of increasing coarseness where the finest and coarsest grid levels are $[-0.5,7.5]c \times [-2,2]c$ and $[-60.5,67.5]c \times [-32,32]c$, respectively. The sub-domain boundaries are set to $[-0.008,3.296]c \times [-0.348,0.208]c$ , which encompasses all the airfoils and the physical limits of flap displacements. 
The grid spacing of the finest domain is $\dx/c=0.002$ and the time step size is $\dt/(c/\velocityscale) = 0.00025$. Note that these discretizations are finer than those considered for the similar airfoil-flap problem considered in the previous section \ref{airfoilproblem}; therefore, a grid convergence study is not performed for this problem. The resulting size of the flow domain is $4000 \times 2000$, or 8 million grid points.

Recall from Sec. \ref{parallelimplementation} that, for parallel implementation, the FFTW-MPI library requires that the domain decomposition of the flow domain be performed along the $y$-direction for 2D problems. For the tandem-airfoil-flap problem, this domain decomposition corresponds to 1D partitioning along the $y$-direction consisting of 2000 grid points. However, the preferred domain partitioning is along the $x$-direction,which has the larger dimension of 4000 grid points. We thus superficially rotate the original computational domain by $90^\circ$ in clockwise direction to obtain a domain of $2000 \times 4000$ points. When displaying the results, the flow-fields are rotated back to the original $4000 \times 2000$ configuration for readability.

%=======================================================================================

\subsubsection{Implementation}
\label{tandemimplementation}

%------------------------------------------------------
\begin{figure}
\centering
\begin{subfigure}[t]{0.49\textwidth}
\centering
\includegraphics[scale=1]{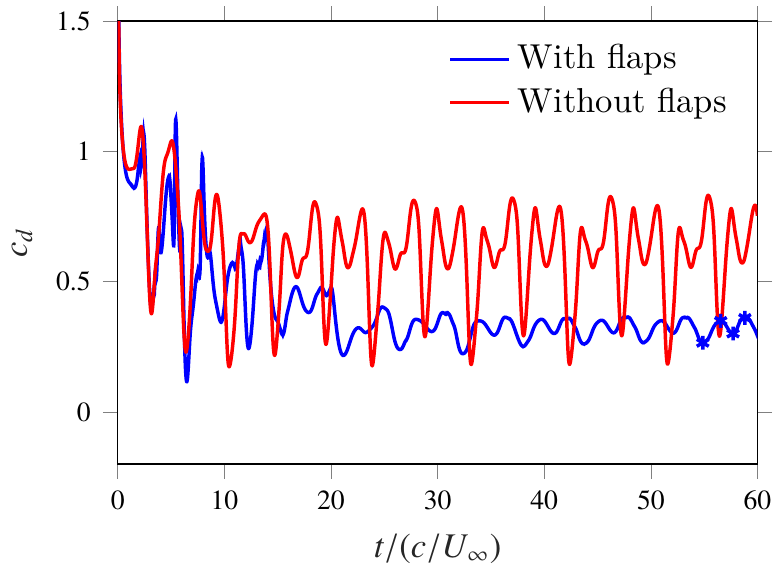}
\caption{Plot of $c_d$ v/s time for the tandem-airfoil systems with and without flaps.}
\label{tandemcd}
\end{subfigure}
\begin{subfigure}[t]{0.49\textwidth}
\centering
\includegraphics[scale=1]{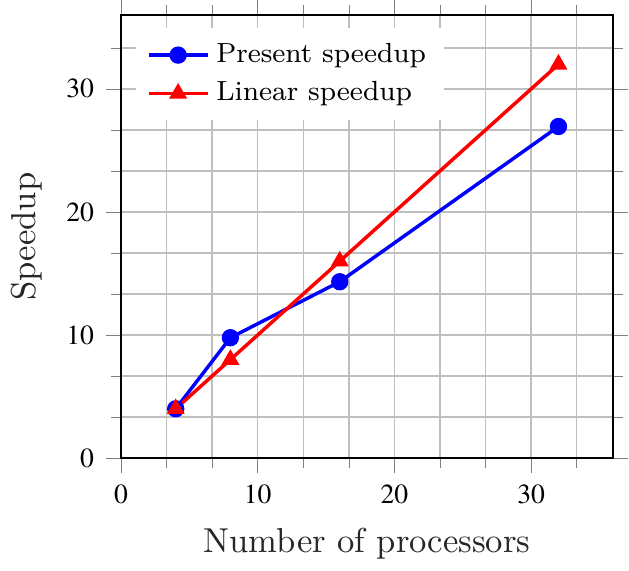}
\caption{Plot of observed and linear speed-up v/s number of processors.}
\label{tandemscaling}
\end{subfigure}
\caption{Comparison of total drag coefficient, $c_d$ (left) and demonstration of favourable strong scaling (right)}
\end{figure}
%------------------------------------------------------
%-------------------------------------------
\begin{figure}
\centering
\begin{subfigure}[t]{1\textwidth}
\centering
\includegraphics[scale=1]{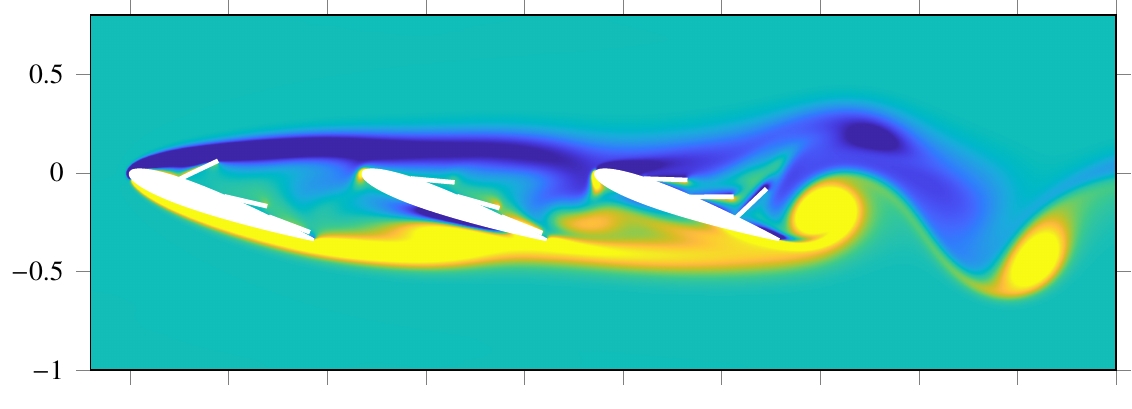}
\caption{$t/T=0$}
\end{subfigure}
\begin{subfigure}[t]{1\textwidth}
\centering
\includegraphics[scale=1]{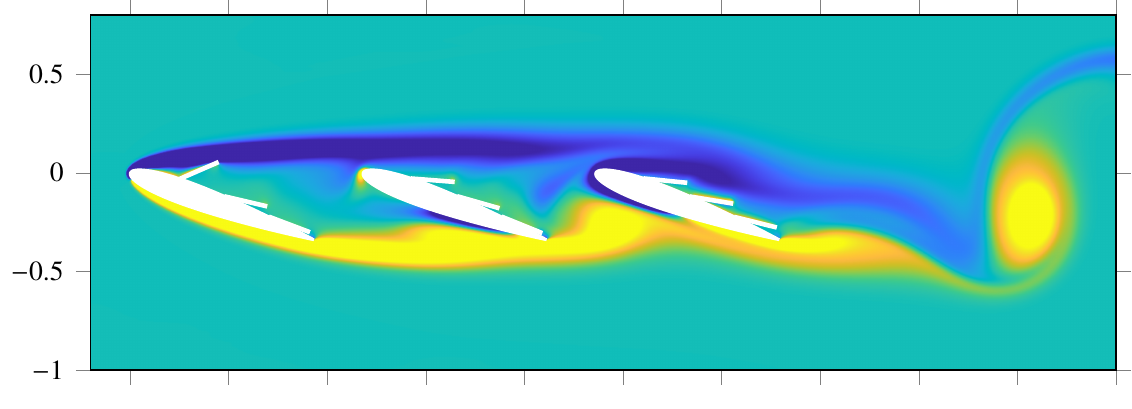}
\caption{$t/T=0.301$}
\end{subfigure}
\centering
\begin{subfigure}[t]{1\textwidth}
\centering
\includegraphics[scale=1]{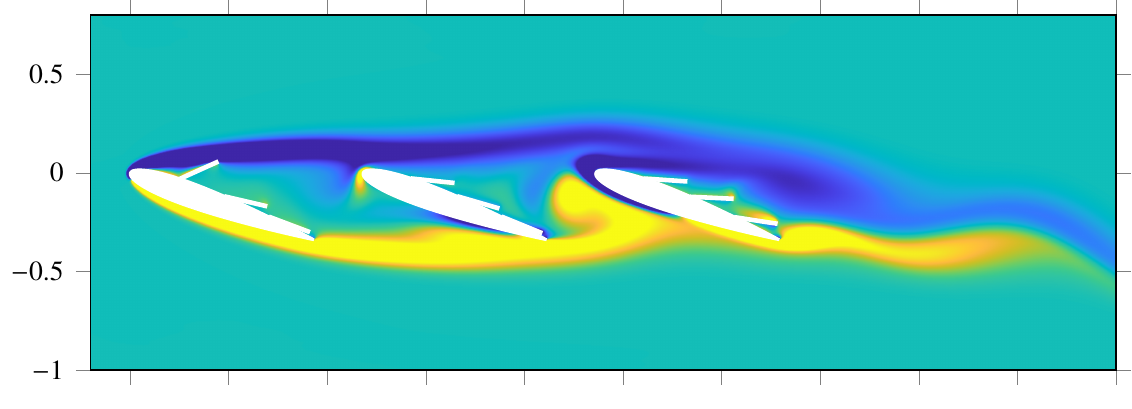}
\caption{$t/T=0.505$}
\end{subfigure}
\begin{subfigure}[t]{1\textwidth}
\centering
\includegraphics[scale=1]{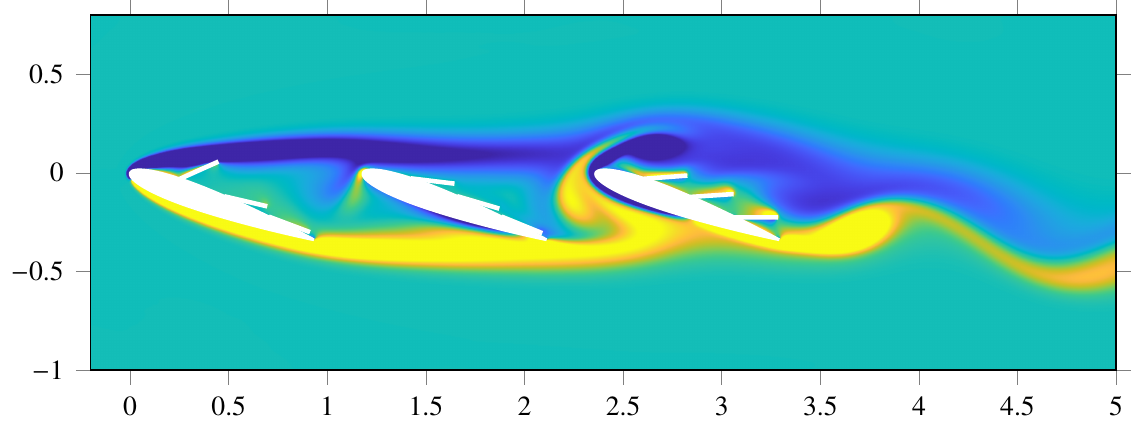}
\caption{$t/T=0.704$}
\end{subfigure}
\caption{Contour plots of vorticity at different time instants corresponding to the blue markers in the $c_d$ plot of Fig. \ref{tandemcd}. Blue and yellow denote regions of counter-clockwise and clockwise vorticity. Here $T$ denotes the time period of limit cycle oscillations.}
\label{tandemsnapshots}
\end{figure}
%------------------------------------------------------

First, the total drag coefficient of the tandem-airfoil-flap system, $c_d = 2F_x/\densityf\velocityscale^2c$, where $F_x$ is the total force on all airfoils and flaps in the $x$-direction, is plotted in Fig. \ref{tandemcd} and compared with the case of the same three airfoils in tandem, but without any flaps. A reduction in mean drag, $\bar{c}_d$, by 46.92\% is observed with respect to the flap-less case, where the mean is evaluated in the limit cycle oscillation regime after initial transients have decayed, $t/(c/\velocityscale)>30$. To indicate the physical mechanisms that enable this drag reduction, four snapshots of vorticity are plotted in Fig. \ref{tandemsnapshots}. These snapshots correspond to two troughs and two peaks of one drag cycle in the limit cycle oscillation regime, indicated by the blue markers on the $c_d$ plot in Fig. \ref{tandemcd}. 
From these figures, we observe that the deployed flaps manipulate the flow to curve around a large ``imaginary body'' that acts as a streamlined connection of the true tandem-airfoil-flap system. Although significant flow separation occurs at the first airfoil, the leading flap deflects the shear layer in the upwards transverse direction, shielding the second and much of the third airfoil from drag-producing vortex interactions. The end result is a net reduction of drag for the collective system. 

Now, we demonstrate favourable strong scaling by evaluating the speedup obtained over the first 1000 time steps ($t/(c/\velocityscale)<0.25$) while increasing the number of processors as $\{4,8,16,32\}$. Typically, the speedup is defined as the ratio of the time taken by one processor to that of $p$ parallel processors. However, due to the large size of the problem, we instead define the speedup with respect to four processors as, 
\begin{equation}
\text{Speedup} = \frac{T_4}{T_p} \times 4
\end{equation}
where $T_p$ is the time taken by $p$ processors. Note that we only take into account the time incurred in the online stage of our algorithm for calculating speedup. The scaling results are displayed in Fig. \ref{tandemscaling} by plotting the speedup versus the number of processors. A plot of linear (ideal) speedup is also provided for reference. A favourable strong scaling efficiency of 84.22\% at $p=32$ processors is observed where efficiency is defined as the ratio of speedup to $p$. Finally, for all the cases considered, a maximum of only two FSI iterations were required per time step.

%=======================================================================================
\pagebreak
\section{Conclusions}
\label{conclusions}

In this manuscript, we have proposed an efficient sub-domain based IB approach that addresses the computational bottleneck encountered in a number of strongly and semi-strongly coupled IB methods, wherein several costly large dimensional systems are solved only for a small number of body variables. In our proposed approach, the fluid-structure coupling operator is constructed on a fixed set of flow sub-domain points instead of time-varying body points, allowing us to precompute a matrix that embeds the large dimensional system before any time advancement is performed. This precomputation process results in all FSI iterations being restricted to small-dimensional systems. As such, the proposed algorithm mimics favorable features of stationary-body IB methods, where the matrix that encodes the interface coupling can be precomputed, while retaining the desirable stability properties of strongly coupled FSI methods. We also formulated a parallel implementation of this sub-domain-based IB algorithm, and demonstrated favorable strong scaling. 

Numerical experiments consisted of two dimensional flow problems involving large body displacements such as flapping of torsionally connected plates and the FSI dynamics of a passively deployable flap on an airfoil. The results obtained from our approach agreed well with those from the previous studies. Regarding computational efficiency, our approach outperformed an implementation of the IB method without the proposed sub-domain approach, delivering speed-ups of up to an order of magnitude for the presented problems. Finally, favorable strong scaling of our parallel implementation was demonstrated on a larger problem consisting of three airfoils in tandem, each equipped with three passively deployable flaps. For all the cases considered, our approach produced a convergent solution in less than three FSI iterations.

In this manuscript, we have developed our sub-domain based IB method on the foundation of the IB method of \citet{goza2017strongly}. However, we emphasize that our formulation can be extended to a wide range of strongly coupled IB methods. Furthermore, we note that although the flow problems considered in this work consisted of a combination of rigid and torsional bodies, the formulation was developed and equally applicable for a more general setting that includes deformable bodies, possibly combined to create more complex structures.

%=======================================================================================
\section{Acknowledgement}

We gratefully acknowledge funding through the National Science Foundation under grant CBET 20-29028. The code for the proposed sub-domain based IB approach used to simulate the problems in this work is open-source and publicly available at \url{https://github-dev.cs.illinois.edu/NUFgroup/IB_parallel}. 
%=======================================================================================
\appendix

\section{Derivation of Eq. \texorpdfstring{\eqref{predictor}--\eqref{corrector}}{}}
\label{appendixa}

This appendix provides the derivation of the fully discretized and block LU factorized equations \eqref{predictor}--\eqref{corrector} from the governing equations \eqref{ns}--\eqref{bcd}. Firstly, the spatially discretized equations of motion for the fluid on a staggered uniform Cartesian grid in the vorticity-streamfunction formulation \cite{colonius2008fast} is given by,
\begin{equation}
\Ct \C \dot{\streamfnc} + \mathcal{N}(\C \streamfnc) = \Ct L\C\streamfnc - \Ct \Etn{}(\bodypointn{})\stressn{}
\end{equation}
where $\mathcal{N}(\cdot)$ is the discretization of the nonlinear advection term.

\begin{figure}
\centering
\includegraphics[scale=1]{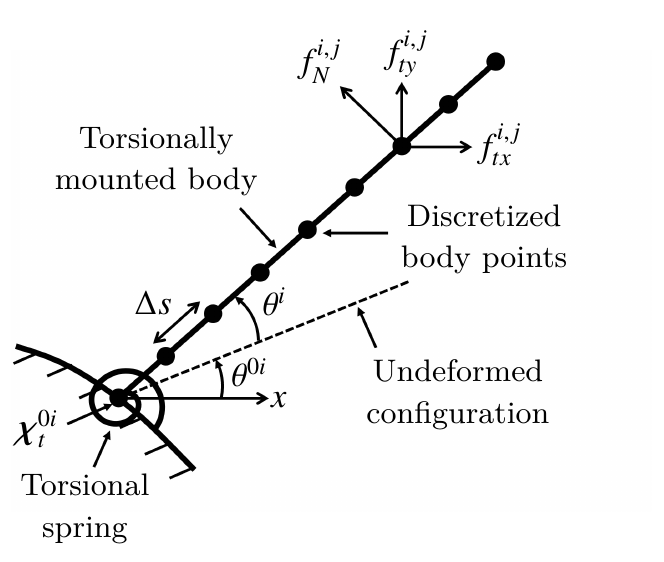}
\caption{Schematic of a torsionally mounted body}
\label{schematictors}
\end{figure}

For the spatial discretization of the equation for the torsionally connected bodies, consider the  schematic of the $i^{th}$ torsional body $\torsbodyi$ with an undeformed (zero stress) angle $\deflectionoi$ from the $x$-axis in Fig. \ref{schematictors}. The normal surface stress, $\stressnormali$, exerted on the body by the fluid is given by,
\begin{equation}
\stressnormali = -\stressn{tx}^i \sin(\deflectionoi + \deflectioni) + \stressn{ty}^i \cos(\deflectionoi + \deflectioni) = \Ri{}{} \stressn{t}^i
\end{equation}
where $\stressn{tx}^i$ and $\stressn{ty}^i$ are the surface stresses in the $x$ and $y$ directions, respectively; $\stressn{t}^i = [\stressn{tx}^i,\stressn{ty}^i]^T$; and $\Ri{}{}$ is a matrix containing two blocks of diagonal matrices aligned column-wise with diagonal entries $ -\sin(\deflectionoi + \deflectioni)$ and $\cos(\deflectionoi + \deflectioni)$ corresponding to $\stressn{tx}^i$ and $\stressn{ty}^i$, respectively. Accordingly, the moment due to surface stress can be discretized as,
\begin{equation}
- \int_{\torsbodyi} (\torsbodypoint-\hinge) \times \stress(\torsbodypoint) d\torsbodypoint \xrightarrow[]{\text{discretize}} \sum_{j=0}^{n_{t}^i} (j\ds) (\Ri{}{} \stressn{t}^i)_j \ds = \Qi{}\Ri{}{}\stressn{t}^i\ds
\end{equation}
where $n_t^i$ is the number of discretized points on $\torsbodyi$ and $\Qi{} = [0,1,\ldots,n_t^i] \ds$. Now, the semi-discretized equations for the tortional body is given by,
\begin{equation}
\inertia \dot{\deflectiondot}^i + \damper \deflectiondoti + \stiffness \deflectioni = \Qi{}\Ri{}{}\stressn{t}^i\ds + g_t^i \quad \quad \text{for} \quad i=1,\ldots,\nbt
\end{equation}
where we define $\deflectiondoti = \dot{\deflection}^i$.

The equation for a deformable body is discretized using a finite element procedure as described in \citet{goza2017strongly}. By expressing the structural variables using a set of compatible shape functions, we write the spatially discretized form of Eq. \eqref{beam} as,
\begin{equation}
\Mdi \dot{\zeta}^i_d + \Rdi(\defbodypointi{}) = \Qdi(g_d^i + \Wdi{}{}(\defbodypointi{})\stressn{d}^i) \quad \quad \text{for} \quad i=1,\ldots,\nbt
\end{equation}
where $\stressn{d}^i = [\stressn{dx}^i,\stressn{dy}^i]^T$ and the specific forms of $\Mdi$, $\Rdi$, $\Qdi$ and $\Wdi{}{}$ containing the shape functions are described in reference \cite{goza2017strongly}. Next, the boundary conditions on all the bodies are discretized as,
\begin{equation}
\En{r}^{i} \C \streamfnc = u_r^i \quad \quad \text{for} \quad i=1,\ldots,\nbr
\end{equation}
\begin{equation}
\En{t}^{i} \C \streamfnc - \Ri{}{T}\Qi{T} \deflectiondoti = 0 \quad \quad \text{for} \quad i=1,\ldots,\nbt
\end{equation}
\begin{equation}
\En{d}^{i} \C \streamfnc - \defbodypointdoti{} = 0 \quad \quad \text{for} \quad i=1,\ldots,\nbd
\end{equation}
%
%where $\En{r}^i$, $\En{t}^i$ and $\En{d}^i$ are the interpolation operators for the rigid, torsional and deformable bodies, respectively and $u_b^i$ is the (optionally) prescribed velocity on the rigid body.
%
Following the time discretization schemes of \citet{goza2017strongly}, the fully discretized equations are written as,
\begin{equation}
\Ct\A\C \streamfnc_{n+1} + \Ct \Etn{n+1}\stressn{n+1} = \fluidrhs
\label{ns3}
\end{equation}
\begin{equation}
\frac{4}{\dt^2}\inertia\deflectioni_{n+1} + \frac{2}{\dt}\damper \deflectioni_{n+1} + \stiffness \deflectioni_{n+1} - \Qi{}\Ri{,n+1}{}{\stresst{,n+1}}\ds = r^{\deflectiondot,i}_{n}  \quad \quad \text{for} \quad i=1,\ldots,\nbt
\end{equation}
\begin{equation}
\frac{2}{\dt}\deflectioni_{n+1} - \deflectiondoti_{n+1} = r_n^{\deflection,i} \quad \quad \text{for} \quad i=1,\ldots,\nbt
\end{equation}
\begin{equation}
\frac{4}{\dt^2}\Mdi\defbodypointi{,n+1} + \Rdi(\defbodypointi{,n+1}) - \Qdi\Wdi{,n+1}{}\stressd{,n+1} = r^{\zeta,i}_{n}  \quad \quad \text{for} \quad i=1,\ldots,\nbd
\label{beam3}
\end{equation}
\begin{equation}
\frac{2}{\dt}\defbodypointi{,n+1} - \defbodypointdoti{,n+1} = r_n^{\bodypointn{},i} \quad \quad \text{for} \quad i=1,\ldots,\nbd
\end{equation}
\begin{equation}
\En{r,n+1}^{i} \C \streamfnc_{n+1} = u_{r,n+1}^i  \quad \quad \text{for} \quad i=1,\ldots,\nbr
\end{equation}
\begin{equation}
\En{t,n+1}^{i} \C \streamfnc_{n+1} - \Ri{,n+1}{T}\Qi{T} \deflectiondoti_{n+1} = 0  \quad \quad \text{for} \quad i=1,\ldots,\nbt
\end{equation}
\begin{equation}
\En{d,n+1}^{i} \C \streamfnc_{n+1} - \defbodypointdoti{,n+1} = 0  \quad \quad \text{for} \quad i=1,\ldots,\nbd
\label{bc3}
\end{equation}
where $\fluidrhs = (\frac{1}{\dt}\Ct\C + \frac{1}{2}\Ct L \C)\streamfnc_n + \frac{3}{2}\Ct \mathcal{N}(\C\streamfnc_n)-\frac{1}{2}\Ct \mathcal{N}(\C\streamfnc_{n-1})$, 
$r^{\deflectiondot,i}_{n} = \inertia\left(\frac{4}{\dt^2}\deflectioni_{n} + \frac{4}{\dt}\deflectiondoti_{n} + \dot{\deflectiondot}^{i}_{n}\right) + \damper\left(\frac{2}{\dt}\deflectioni_{n} + \deflectiondoti_{n}\right) + g_t^i$,
$r_n^{\deflection,i} = \deflectiondoti_{n} + \frac{2}{\dt}\deflectioni_{n}$,
$r^{\zeta,i}_{n} = \Mdi\left(\frac{4}{\dt^2}\defbodypointi{,n} + \frac{4}{\dt}\defbodypointdoti{,n} + \dot{\zeta}^{i}_{d,n}\right) + \Qdi g_d^i$ and
$r_n^{\bodypointn{},i} = \defbodypointdoti{,n} + \frac{2}{\dt}\defbodypointi{,n}$. 
Following \citet{goza2017strongly}, an iterative procedure is introduced to solve the above system of equations. A guess at iteration $(k)$ is used to compute a new guess at $k+1$ by defining, $\psi_{n+1}^{i(k+1)} = \psi_{n+1}^{i(k)} + \Delta\psi^{i}$ where $\psi=\{\deflection,\deflectiondot,\chi_d,\zeta_d\}$ and $\Delta\psi^i$ is assumed to be small. On substituting this decomposition into \eqref{ns3}-\eqref{bc3} and retaining first order terms in the increments and $\dt$, we get,
\begin{equation}
%\begin{bmatrix}
%\Ct\A\C & 0 & 0 & 0 & 0 & \Ct\Etnk{n+1}{k} \\
%0 & 0 & \J & 0 & 0 & (0 \ \ -\Q{(k)}\R{,n+1}{(k)}\ds \ \ 0) \\
%0 & -I_t & \frac{2}{\dt}I_t & 0 & 0 & 0 \\
%0 & 0 & 0 & 0 & \Jd & (0 \ \ 0 \ \ -\Qd{(k)}\Wd{,n+1}{(k)}) \\
%0 & 0 & 0 & -I_d & \frac{2}{\dt}I_d & 0 \\
%\Enk{n+1}{k}\C & 
%%
%\begin{pmatrix}
%0 \\
%-\R{,n+1}{(k)T}\Q{(k)T} \\
%0
%\end{pmatrix} & 0 & 
%%
%\begin{pmatrix}
%0 \\
%0 \\
%-I_d
%\end{pmatrix} & 0 & 0 \\
%\end{bmatrix}
%
\begin{bmatrix}
\Ct\A\C & 0 & 0 & 0 & 0 & \Ct\Etnk{n+1}{k} \\
0 & 0 & \J & 0 & 0 & \Snk{n+1}{k}\ds \\
0 & -I_t & \frac{2}{\dt}I_t & 0 & 0 & 0 \\
0 & 0 & 0 & 0 & \Jd & \Sdnk{n+1}{k} \\
0 & 0 & 0 & -I_d & \frac{2}{\dt}I_d & 0 \\
\Enk{n+1}{k}\C & \Stnk{n+1}{k} & 0 & \hat{I}_d^T & 0 & 0 \\
\end{bmatrix}
\begin{bmatrix}
\streamfnc_{n+1} \\
\deflectiondotincrease\\
\deflectionincrease\\
\defbodypointdotincrease\\
\defbodypointincrease\\
\stressnk{n+1}{k}
\end{bmatrix} = 
\begin{bmatrix}
\fluidrhs + \mathcal{O}(\dt) \\ 
r^{\deflectiondot}_{n} -\J\deflectionnk{n+1}{k} + \mathcal{O}(\dt)\\
r_n^{\deflection} - \frac{2}{\dt}\deflectionnk{n+1}{k} +\deflectiondotnk{n+1}{k}\\ 
r^{\zeta}_{n} -\Jd\defbodypointnk{n+1}{k} + \mathcal{O}(\dt)\\
r_n^{\bodypointn{}} - \frac{2}{\dt}\defbodypointnk{n+1}{k} +\defbodypointdotnk{n+1}{k}\\ 
U_{b,n+1}^{(k)T} + \mathcal{O}(\dt)
%\begin{pmatrix}
%u_{b,n+1} \\
%\R{,n+1}{(k)T}\Q{(k)} \deflectiondotnk{n+1}{k}
%\end{pmatrix} + \mathcal{O}(\dt)
\end{bmatrix} :=
\begin{bmatrix}
\fluidrhs\\ 
\deflectiondotrhs{k} \\
\deflectionrhs{k} \\ 
\defbodypointdotrhs{k} \\
\defbodypointrhs{k} \\ 
\bcrhs{k}
\end{bmatrix}
\label{bigsystem}
\end{equation}
Here, we have aggregated all the individual $\Delta\psi^i$ into a vector $\Delta \psi$ where $\psi=\{\deflection,\deflectiondot,\chi_d,\zeta_d\}$ and for the right-hand side terms. $I_t$ and $I_d$ are identity operators of compatible sizes for the torsional and deformable bodies, respectively; $\J$ is a square diagonal operator of size $\nbt$ with diagonal elements ${\J}_{(i,i)} = \frac{4}{\dt^2}\inertia + \frac{2}{\dt}\damper + \stiffness$; and $\Jd$ is a square block diagonal operator having $\nbd$ blocks where the $i^{th}$ diagonal block is given by $J_{d(i,i)}^{(k)} = \frac{4}{\dt^2}\Mdi + K_{d}^{i(k)}$ where $K_{d}^{i(k)} = d\Rdi/\chi|_{\chi=\chi_{d,n+1}^{i(k)}}$. The remaining operators are defined as 
$\Snk{n+1}{k} = [0,\ -\Q{(k)}\R{,n+1}{(k)}\ds,\ 0]$,
$\Sdnk{n+1}{k} = [0,\ 0,\ -\Qd{(k)}\Wd{,n+1}{(k)}]$, 
$\hat{I}_d = [0,\ 0, \ I_d]$, 
$U_{b,n+1}^{(k)} = [u_{r,n+1},\ \R{,n+1}{(k)T}\Q{(k)} \deflectiondotnk{n+1}{k},\ \defbodypointdotnk{n+1}{k}]$, where 
$\R{}{}$ , $\Q{}$, $\Wd{}{}$ and $\Qd{}$ are block diagonal operators with entries $\Ri{}{}$, $\Qi{}$, $\Wdi{}{}$ and $\Qdi{}$, respectively. On performing a block LU decomposition of Eq. \eqref{bigsystem}, we get the final system of equations given in Eq. \eqref{predictor}--\eqref{corrector}.
%=======================================================================================

\bibliographystyle{elsarticle-num-names}
\bibliography{references.bib}

\end{document}